\documentclass[prd,twocolumn,10pt,aps,superscriptaddress,longbibliography,nofootinbib]{revtex4-2}
\pdfoutput=1

\usepackage{subfig,tabularx,afterpage,xifthen}
\usepackage{amsmath,amssymb,amsfonts,mathtools}
\usepackage{appendix}
\usepackage{graphicx}
\usepackage{color,xcolor}
\usepackage{cancel}
\usepackage{hyperref}
\hypersetup{
    colorlinks=true,
    linkcolor=black,
    citecolor=black,      
    urlcolor=black,
    }

\begin{document}

\title{Gravitational waves and tadpole resummation: Efficient and easy convergence of finite temperature QFT}
\author{David Curtin}
\email{dcurtin@physics.utoronto.ca}
\author{Jyotirmoy Roy}
\email{jro1@physics.utoronto.ca}
\affiliation{Department of Physics, University of Toronto, Toronto, Ontario M5S 1A7, Canada}
\author{Graham White}
\email{G.A.White@soton.ac.uk}
\affiliation{Kavli IPMU (WPI), UTIAS, The University of Tokyo, Kashiwa, Chiba 277-8583, Japan}

\begin{abstract}
We demonstrate analytically and numerically that  ``optimized partial dressing'' (OPD) thermal mass resummation, which uses  gap equation solutions inserted into the tadpole, efficiently tames finite-temperature perturbation theory calculations of the effective thermal potential, without necessitating use of the high-temperature approximation. An analytical estimate of the scale dependence for OPD resummation, standard Parwani resummation (Daisy resummation), and dimensional reduction shows that OPD has similar scale dependence to dimensional reduction, greatly improving Parwani resummation. We also elucidate how to construct and solve the gap equation for realistic numerical calculations, and demonstrate OPD's improved accuracy for a toy scalar model. OPD's improved accuracy is most physically significant when the high-temperature approximation breaks down, rendering dimensional reduction unusable and Parwani resummation highly inaccurate, with the latter  underestimating the maximal gravitational wave amplitude for the model by 2 orders of magnitude compared to OPD. Our work highlights the need to bring theoretical uncertainties under control even when analyzing broad features of a model. Given the simplicity of the OPD compared to two-loop dimensional reduction, as well as the ease with which this scheme handles departures from the high-temperature expansion, we argue this scheme has great potential in analyzing the parameter space of realistic beyond the Standard Model models.
\end{abstract}

\maketitle

\section{Introduction}

A key aim of next generation experiments is to reveal the nature of cosmological electroweak symmetry breaking. It is expected that future colliders could definitively rule out or confirm a strong first-order electroweak phase transition \cite{Curtin:2014jma, Kotwal:2016tex,Ramsey-Musolf:2019lsf,Papaefstathiou:2020iag}. This departure from thermal equilibrium could supply one of the necessary ingredients for baryogenesis, explaining why there is more matter than antimatter \cite{Trodden:1998qg,Cline:2006ts,Morrissey:2012db,White:2016nbo,Garbrecht:2018mrp}, and produce a stochastic gravitational wave background \cite{Ramsey-Musolf:2019lsf,Profumo:2007wc,Delaunay:2007wb,Huang:2016cjm,Chala:2018ari,Croon:2020cgk,Grojean:2006bp,Alves:2018jsw,Alves:2020bpi,Vaskonen:2016yiu,Dorsch:2016nrg,Chao:2017vrq,Wang:2019pet,Demidov:2017lzf,Ahriche:2018rao,Huang:2017rzf,Mohamadnejad:2019vzg,Baldes:2018nel,Huang:2018aja,Ellis:2019flb, Alves:2018oct, Alves:2019igs,Cline:2021iff,Chao:2021xqv,Liu:2021mhn,Zhang:2021alu,Cai:2022bcf} that can be measured by LISA \cite{Caprini:2019egz} and other experiments in the mHz to kHz frequency range \cite{Punturo:2010zz,Yagi:2011wg,AEDGE:2019nxb,Hild:2010id,Sesana:2019vho,Theia:2017xtk}. Besides this, strong phase transitions can appear in hidden sectors \cite{Schwaller:2015tja,Baldes:2018emh,Breitbach:2018ddu,Croon:2018erz,Hall:2019ank,Baldes:2017rcu,Geller:2018mwu,Croon:2019rqu,Hall:2019rld,Chao:2020adk,Dent:2022bcd}, symmetry breaking chains in grand unified gauge theories \cite{Hashino:2018zsi,Huang:2017laj,Croon:2018kqn,Brdar:2019fur,Huang:2020bbe}, conformal extensions of standard model \cite{Prokopec:2018tnq,Kierkla:2022odc} and any number of other motivated scenarios \cite{Caldwell:2022qsj}. In all cases, a proper treatment of perturbation theory is needed in order to make the theory predictive. 

Unfortunately, 4D perturbation theory converges slowly at finite temperature (for a recent discussion of the convergence see Ref \cite{Schicho:2022wty}). Naively, the distribution function diverges for long wavelength modes \cite{Linde:1980ts}. This issue is delayed if the theory is resummed such that the strongly coupled, long wavelength behavior is screened by hard thermal loops \cite{Parwani:1991gq,Arnold:1992fb,Arnold:1992rz}. However, even after such a ``Daisy resummation'' the perturbation theory converges slowly and as a result a one-loop calculation can predict a peak gravitational wave amplitude that varies by multiple orders of magnitude \cite{Gould:2021oba,Croon:2020cgk} for reasonable choices of the renormalization scale. 

A known solution is to integrate out heavy Matsubara modes, which results in a 3D effective field theory. If one defines such a theory, both in its matching and the expansion in the dimensionally reduced theory, at next-to-leading order (NLO) in the appropriate coupling constant,\footnote{There is something of an inconsistency in the literature as to what prescription corresponds to what order. We will use the convention in this paper that NLO is when a calculation is performed accurately in dimensional reduction to $O(g^4)$, such that performing the resummation at NLO and calculating the effective potential at NLO within the effective theory corresponds to correctly defining the theory up to $O(g^4)$.} the uncertainty in the gravitational wave spectrum can in some cases be reduced to the percent level \cite{Croon:2020cgk}.  
Further, recent work has demonstrated that such perturbative calculations in the dimensionally reduced theory reproduce 3D lattice results very well, indicating that the infrared problems at higher loops may not be numerically important \cite{Ekstedt:2022zro,Gould:2022ran}.\footnote{Two caveats deserve to be mentioned. First the timescale of the phase transition had strong uncertainties in the NLO dimensionally reduced theory. Second, the power counting used is slightly different to what has been used before, expanding in the ratio of the quartic coupling to the gauge coupling, $x\sim\lambda/g^2$, rather than the gauge coupling $g^2$.} The difficulty in dimensional reduction is its tractability. For even the Standard Model effective theory, defining the effective potential at noticeably better accuracy than (convenient) standard 4D methods requires the calculation of $O(10^2)$ diagrams at finite temperature. While a recent package adds automation to this process \cite{Ekstedt:2022bff}, one still has to monitor whether it is appropriate to have multiple dynamical fields and whether to use the soft or ultrasoft potential for different regions of the parameter space. Further, it is difficult to go beyond the high-temperature expansion in dimensional reduction as there is no longer a hierarchy between the lightest Matsubara mode and the soft scale to justify an effective field theory.
This is an important limitation, since strong phase transitions require sizable couplings to the scalar undergoing the transition, which in turn leads to large field-space dependencies for particle masses and hence breakdowns in the high-temperature expansion. 
Therefore, it is strongly motivated to find more convenient 4D calculational methods that can achieve similar levels of accuracy. This would be of significant utility in examining the large theory space of beyond the Standard Model (BSM) scenarios with strong phase transitions, electroweak or hidden.

One candidate for such a resummation method, called ``partial dressing'', was first developed three decades ago~\cite{Boyd:1993tz} (see also \cite{Espinosa:1992gq,Espinosa:1992kf,Dine:1992wr,Boyd:1992xn}) in the context of simple a $\phi^4 $ toy model. 
It represented a simple analytical way of resumming the most important higher-order corrections to the thermal propagator without double counting.
More recently, this method was revisited and adapted for numerical application to the full SM with an additional scalar~\cite{Curtin:2016urg}. Referred to as ``optimized partial dressing'' (OPD), this method was also shown to be easily applied outside of the high-temperature approximation, promising a more accurate treatment of strong phase transitions in this important regime. However, significant further work is required before OPD could become a gold standard for finite-temperature calculations. A systematic and rigorous study of the scheme's convergence and validity is outstanding. 
There are open questions on the detailed analytical construction of the gap equation as a function of scalar vacuum expectation value (VEV), and the method of its numerical solution.\footnote{In fact, the ``optimized'' part in OPD refers to the realization that the full finite-temperature thermal functions can be used in the potential while maintaining use of the high-temperature approximation in the gap equations to make their solution tractable, as well as the particular way mass derivatives are included in the gap equation away from the origin to find sensible $M^2(\phi)$ trajectories. The latter part will be refined in this work.}
Finally, gauge bosons were not yet consistently included in the system of gap equations, and OPD's relationship to renomalization group (RG) improvement of the effective potential is unclear.

The OPD scheme consists of two steps. The dominant contributions  from many higher-order diagrams are conveniently included in a gap equation for the thermal mass. The convenience arises from the fact that the diagrams do not need to be evaluated analytically, one merely needs to solve the gap equation, analytically in some approximations but generally numerically. It is also straight forward to handle cases where the high-temperature expansion breaks down as one can merely use the full one-loop thermal functions within the gap equation. The second step involves inserting the full thermal mass into the tadpole, rather than the full one-loop effective potential as it was rigorously demonstrated that this is the way to prevent double counting of higher order diagrams \cite{Boyd:1993tz}. Finally, missing diagrams are easily identified and can be added by hand. A point to note is that while resummed two-loop potentials have been evaluated without resorting to high-temperature expansion in \cite{Laine:2017hdk}, the current work differs from the former since it includes subleading diagrams via  the correct OPD resummation procedure, in addition to dealing with full thermal functions.
Apart from convenience and handling cases where the high-temperature expansion breaks down, it is interesting to note that a (nonstandard) dimensional reduction calculation \emph{with gap resummation} finds a critical Higgs mass at which electroweak symmetry begins to become first order \cite{Buchmuller:1994qy} as well as a critical end point in QCD at finite density \cite{Gao:2021nwz}, unlike standard dimensional reduction or 4D perturbative calculations. This suggests solving the gap equation may even be the missing ingredient to accurately characterize the SM phase diagram within a perturbative treatment.

In this work, we carefully study OPD in the context of two simple test models---a single-field $\phi^4$ theory, and a two-field $\phi^4$ scalar field theory with a discrete $Z_2 \times Z_2$ symmetry, only one of which is broken. We first analytically show that an RG-improved effective OPD thermal potential is parametrically as accurate as the RG-improved potential calculated using two-loop dimensional reduction, which formally motivates fully developing the OPD calculation in this and subsequent works. 
We then clarify construction of the gap equation and the method of its numerical solution, demonstrating that  the corresponding effective potential improves on conventional Daisy resummation schemes~\cite{Parwani:1991gq,Arnold:1992fb,Arnold:1992rz} in accuracy and precision. Specifically, we compute the thermal potential near the critical temperature, compute the nucleation temperature and strength of the gravitational wave signal, and evaluate the  dependence of these predictions on the renormalization scale to assess the degree of convergence. Unlike dimensional reduction, OPD is applicable in physical regimes  where the high-temperature approximation does not apply, but when applied to scenarios where this approximation is valid, OPD yields results that are very close to the two-loop dimensional reduction calculation. This is in line with  expectations derived from our analytical results.
\footnote{Note that for all our numerical studies in this paper, we do \emph{not} RG-improve the effective potentials. While our analytical results indicate that RG-improved one-loop OPD will yield a similar accuracy as RG-improved two-loop dimensional reduction, it is not yet clear how to RG-improve a numerical OPD calculation in practice; previous calculations of RG-improved perturbative thermal potentials like~\cite{Gould:2021oba} implicitly assume the high-temperature approximation, and eliminating reliance on this approximation is one of the most important reasons to study OPD. We will address this in an upcoming publication.}

Our simple test calculation can be seen as either obtaining results for a particularly simple hidden sector that undergoes a strong phase transition, or as a toy model for SM extensions with extra scalars that achieve a strong electroweak phase transition. At any rate, obtaining careful results for this simple scenario allowed us to improve and more completely understand several analytical and numerical aspects of the OPD method, setting the stage for future work to further develop OPD to include RG-improvement, the contributions of fermions and gauge bosons, momentum dependent contributions to the gap equations and nonrenormalizable operators in the Lagrangian.

The structure of this paper is as follows. In Secs.~\ref{Sec:Resumm_methods} and \ref{s.gw}, we review the different resummation schemes in perturbative finite temperature field theory and computation of the gravitational wave signal. In Sec.~\ref{Sec:Analytics}, we study analytically the scale dependence of physical predictions in the $\phi^4$ theory and the two scalar field theory. The numerical implementation and analysis, including detailed construction of the gap equation away from the origin, is discussed in Sec.~\ref{Sec:Numerics}. We finally conclude our paper with a discussion in Sec.~\ref{Sec:Conclusion}.

\section{Perturbation theory at finite temperature and resummation methods}
\label{Sec:Resumm_methods}
In this section we briefly review finite-temperature perturbation theory and the different resummation schemes we compare in this paper.

The form of gravitational wave spectra generated by a cosmological first-order phase transition ends up being quite sensitive to the precise description of the effective potential.
For example, in the case of the Standard Model effective field theory or simple SM extensions with scalars, in regions where the theory has a strong phase transition the uncertainty in the peak gravitational wave amplitude can be multiple orders of magnitude~\cite{Croon:2020cgk,Gould:2021oba,Gould:2021oba}.

The large uncertainty can be understood in two steps. First, finite-temperature two-loop (and sometimes higher-order) contributions can be similar in size to zero-temperature one-loop pieces. Second, any modest uncertainty in the temperature at which a phase transition occurs is amplified substantially in the prediction of gravitational wave observables. To see the second point, consider that the peak amplitude of a gravitational wave spectrum from a cosmic phase transition (see Sec.~\ref{s.gw}) scales as
\begin{equation}
    \Omega _{\rm GW } \sim \alpha ^2 (\beta/H_\ast )^{-2}
\end{equation}
where $\alpha \sim \Delta V/ T^4 $ is the change in the trace anomaly and $\beta/ H_\ast$ is the inverse timescale of the transition \cite{Ellis:2018mja,Ellis:2019oqb,Guo:2020grp,Gowling:2022pzb}. The difference in pressure between the two phases, $\Delta V$, tends to decrease with temperature and, from dimensional analysis, may typically scale quartically with the inverse temperature.  The inverse timescale typically scales exponentially, so if the uncertainty in $T$ is not too large one can take this as scaling linearly with the inverse temperature. Altogether, this accounts for a remarkable scaling relationship for the peak amplitude
\begin{equation}
    \Omega _{\rm GW } \sim T^{-18}
\end{equation}
such that an $O(25\% )$ uncertainty in the percolation temperature can lead to two orders of magnitude uncertainty in gravitational wave observables.

The parameters responsible for producing a gravitational wave amplitude can independently constrained by measurements at future colliders. However, if the gravitational wave observables are to provide meaningful constraints on the parameters, the theoretical uncertainties need to be brought under control. In fact, as we will later see, even broad questions about a model such as ``what is the largest possible gravitational wave amplitude consistent with this model across its parameter space?'' can vary by orders of magnitude comparing crude and more sophisticated calculation techniques.

We follow standard procedure in attempting to measure the importance of neglected higher-order terms by measuring the renormalization scale dependence of various observables \cite{Croon:2020cgk}. In doing so one needs to make a somewhat arbitrary choice of what range of values for the renormalization scale one should use in the loop calculation. Even for the generous range of scale parameter variation  considered in \cite{Gould:2021oba}, it is possible for the next-to-leading order predictions to lie outside the uncertainty bands of the leading order prediction.
However, the scale variation still gives a good parametric estimate of the size of the missing NLO terms. The conventional resummation methods that result in these large theoretical uncertainties were developed by Parwani~\cite{Parwani:1991gq} as well as Arnold and Espinosa \cite{Arnold:1992fb,Arnold:1992rz}. In these schemes, the long distance behavior is screened by including a thermal mass term, $\Pi$, such that the resummed potential has the form
\begin{eqnarray}
\label{e.Msqsub}
    V_{{\rm 1-loop}}(M^2) \to     V_{{\rm 1-loop}}(M^2+\Pi) \ .
\end{eqnarray}
The thermal mass and the finite-temperature one-loop potential are readily calculated using standard methods.\footnote{See e.g.~\cite{quiros1998finite}, or~\cite{Curtin:2016urg} for a brief summary using the same notation and conventions as our analysis.} It is useful to recall the form of $\Pi$ in the high-temperature expansion for a $\phi^4$ theory with quartic coupling $g$:
\begin{equation}
    \label{e.pi}
    \Pi \sim g^2 T^2 -\frac{g^2 T M}{8 \pi} - \frac{g^2 M^2}{32 \pi^2} \log \frac{\mu^2}{T^2} + \ldots
\end{equation}
Both of these traditional resummation schemes only resum the leading $\mathcal{O}(T^2)$ term in the above thermal mass expansion, but they differ in some details. Arnold-Espinosa adds a ring term to the effective potential, which can be found by performing the substitution in Eq.~(\ref{e.Msqsub}) in the high-temperature-expanded thermal potential only.
Parwani, on the other hand, substitutes $M^2+\Pi$ into the full effective potential, including the zero-temperature Coleman-Weinberg potential. 
For reasons that are beyond the scope of this quick discussion, the field tended to use the Arnold-Espinosa scheme, but the Parwani scheme actually has better scale dependence, since the thermal mass contribution in the zero-temperature Coleman-Weinberg piece induces a partial cancellation with the finite-temperature potential piece. We therefore use the Parwani scheme [referred to as truncated full dressing (TFD) in~\cite{Curtin:2016urg}] to minimize the scale dependence of traditional Daisy resummation, leading to the most conservative assessment of the benefits of OPD or DR.

The Parwani resummation scheme correctly reorganizes the theory such that all pieces up to third order in the $SU(2)_L$ gauge coupling or larger are included in the potential. However, $g_2$ is reasonably large in the Standard Model, and terms at least in the next order $\mathcal{O}(g_2^4)$ are needed to bring uncertainties in the gravitational wave amplitude under control. Organizing perturbation theory with Parwani resummation to include all terms of $O(g_2^4)$ is highly nontrivial, as naive methods lead to double counting.

To improve the accuracy of finite temperature perturbation theory, dimensional reduction at next to next-to-leading order appears to provide a recipe, reducing theoretical uncertainties to the percent level \cite{Croon:2020cgk}. Dimensional reduction relies on the observation that in imaginary time, the quantum field theory is identical to a three-dimensional theory with a compactified time dimension whose size is determined by the temperature \cite{Kajantie:1995dw}. One can then integrate out the heavy Kaluza-Klein like modes, known as Matsubara modes, leaving an effective field theory in three dimensions. If a scale hierarchy persists between the remaining states, typically a soft and ultrasoft scale of order $g T$ and $g^2 T/\pi$ respectively, the soft states can be integrated out leaving behind a simpler effective field theory again. 

Dimensional reduction naturally incorporates resummation through the matching relations between the four dimensional theory and the effective dimensionally reduced theory. Calculating the relevant self-energy diagrams in these matching relations can be done at multiloop levels of accuracy in addition to defining the effective potential at multiple loops within the effective theory without double counting. The whole process makes it straightforward albeit work intensive to organize perturbation theory into powers of an effective coupling constant (or a ratio of constants as argued for in ref. \cite{Gould:2022ran}). At next-to-leading order [or $O(g^4)$] the theoretical uncertainties in the gravitational wave amplitude for the Standard Model effective field theory are effectively under control. The drawback of dimensional reduction is both the enormous practical difficulty of the scheme---over a hundred diagrams are required to see noticeable improvement compared with conventional methods---and the fact that it conventionally assumes the validity of the high temperature expansion.

Another method of resummation---partial dressing or gap resummation calculates many higher-order diagrams numerically by solving a gap equation, which can be pictorially represented as follows:
\begin{widetext}
\begin{figure}[htbp]
    \centering
    \includegraphics[width=0.8\textwidth]{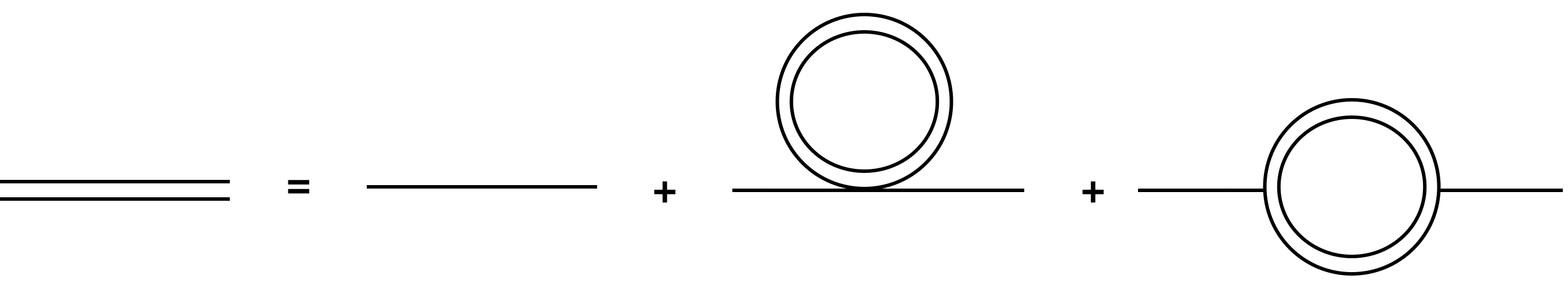}
    \label{fig:my_label}
\end{figure}
\end{widetext}
where the double lines represent the resummed mass and the single line the tree-level mass. It was demonstrated three decades ago~\cite{Boyd:1993tz} that in order to avoid double counting of higher-order diagrams, one needs to insert the resummed mass into the tadpole and integrate,
\begin{equation}
V=\int d \phi \left[ \frac{\partial V_{\text{1-loop}}\left(M^2  \right)}{\partial \phi} +\frac{\partial V_{\text{sun}}\left( M^2 \right)}{\partial \phi} \right]
\end{equation}
where the first term, $V_{\text{1-loop}}$, is the zero plus finite-temperature one-loop potential and the second term is the finite-temperature potential due to two loop sunsets. This term is neglected in the solution to the gap equation.
The method can also be applied when the high-temperature regime breaks down, in particular by keeping the high-temperature approximation in the gap equation but using full thermal integrals in the potential~\cite{Curtin:2016urg} (referred to as OPD), and is much easier to use than dimensional reduction as the gap equation only requires one to define the one-loop effective potential at finite temperature. It is therefore highly attractive if it can be demonstrated to provide substantial improvement over conventional methods, i.e. Parwani resummation.
In this work, we seek to ascertain systematically for the first time whether gap resummation, or OPD, performs better than Parwani resummation and is comparable to dimensional reduction at next-to-leading order for the purposes of obtaining gravitational wave predictions.
We also clarify how the gap equation should be constructed and numerically solved away from the origin. 

\section{Review of gravitational wave signal calculation}
\label{s.gw}

Any uncertainties in the prediction of the thermal potential, arising from the slow convergence of perturbation theory at finite temperature, become amplified when calculating the gravitational wave observables. In this section we will review the calculation of the gravitational wave peak amplitude due to the acoustic contribution calculated using the sound shell model \cite{Hindmarsh:2013xza,Hindmarsh:2015qta,Hindmarsh:2017gnf,Hindmarsh:2019phv}. We will later consider the two scalar field model and assume that the Standard Model particle content is coupled strongly enough to maintain kinetic equilibrium with our two new scalar fields but weakly enough that we can ignore their effect on the gravitational wave phenomenology. In the sound shell model, the spectrum is completely determined by the temperature of the transition, the fluid velocity, the mean bubble separation, the life time of the sound waves and the fraction of energy released that becomes converted to sound waves. To a good approximation, these quantities can be related to four macroscopic parameters - the transition temperature, the bubble wall velocity, the trace anomaly normalized by the critical density
\begin{equation}
    \alpha = \frac{\Delta V - T d \Delta V/dT}{\rho _c}
\end{equation}
and the inverse lifetime of the transition
\begin{eqnarray}
    \frac{\beta }{H_\ast} = T \frac{d (S_E/T)}{dT} \ .
\end{eqnarray}

A more diligent calculation irons out errors that are orthogonal to our analysis \cite{Guo:2021qcq} so we ignore them in this work. It should be noted, however that in the sound shell model, the four thermal parameters result in two observables, the peak amplitude and frequency. More careful simulations yield a more optimistic picture where all four thermal parameters can be extracted from the precise shape of the power spectrum \cite{Gowling:2021gcy}, but conservative analyses claim fits to three parameters~ \cite{Giese:2021dnw}.
In any case, the sound shell model predicts a spectral shape, of the form
\begin{eqnarray}
	S_{\rm SW} (f) = \left( \frac{f}{f_{\rm SW}}\right)^3 \left[ \frac{7}{4 + 3(f/f_{\rm SW})^2}\right]^{7/2} ,
\end{eqnarray}
with a peak frequency
\begin {eqnarray}
f_{\rm SW} &=& 1.9 \times 10^{-5} \frac{1}{v_w} \left( \frac{\beta}{H_n}\right) \nonumber \\
&& \times \left( \frac{T_n}{100 \: \rm GeV}\right) \left( \frac{g_{*}}{100}\right)^{1/6} \rm Hz,
\end{eqnarray}
and a peak amplitude \cite{Hindmarsh:2013xza,Hindmarsh:2015qta,Hindmarsh:2017gnf,Hindmarsh:2019phv,Guo:2020grp}
\begin{eqnarray}
h^2 \Omega_{\rm GW} &=& 8.5 \times 10^{-6} \left( \frac{100}{g_*}\right)^{1/3} \nonumber \\ 
&& \times \kappa^2 \left( \frac{H_*}{\beta}\right) v_{w} \Upsilon \left(\bar{U}_{f,\rm max},R_* \right) \ ,
\label{eq:omega_suppressed} 
\end{eqnarray}
where $v_w$ is the bubble wall velocity and $g_*$ counts the number of relativistic degrees of freedom in the plasma at the time of transition. For the efficiency we assume a relativistic bubble wall velocity which means we can, to a good approximation, relate the trace anomaly to the kinetic energy fraction
\begin{align}
    \kappa &\simeq \frac{\alpha}{0.73 + 0.083 \sqrt{\alpha} + \alpha}
\end{align}
and the suppression factor from the finite lifetime of the source is
\begin{eqnarray}
    \Upsilon = 1 - \frac{1}{\sqrt{1 - 2 \tau_{\rm sw} H_s}},
    \label{eq:suppression-upsilon}
\end{eqnarray}
where $\tau_{\rm sw} = R_*/\bar{U}_f$ with the fluid velocity and the mean bubble separation having the form $U_f^2\sim\frac{3}{4} \kappa \alpha $ and $R_\ast = (8 \pi)^{1/3}v_w/\beta$, respectively.
Finally, to derive the percolation temperature $T_p$ one has to solve the equation,
\begin{eqnarray}
    \frac{S_3(T_p)}{T_p} &=& 131 - \log(A/T^4) - 4 \log\left( \frac{T}{100 \: \rm GeV}\right) \nonumber \\
    && - 4 \log\left( \frac{\beta(T)/H}{100}\right) + 3 \log(v_w).
    \label{eq:LISA_Tp}
\end{eqnarray}

\section{Analytic comparison of resummation methods for \texorpdfstring{$\phi^4$}{} theories}

\label{s.analyticcomparison}

Considering single-$\phi^4$ and two-$\phi^4$ theories with $\mathbb{Z}_2$ symmetry and working in the $\overline{\rm MS}$ renormalization scheme, we now follow a similar procedure as~\cite{Gould:2021oba} in the high-temperature limit to analytically show that the effective potential computed in  an RG-improved one-loop OPD calculation has parametrically similar scale dependence to the RG-improved two-loop dimensional reduction calculation, greatly improving on  standard Parwani resummation. This motivates full development of the OPD calculation, including the progress we make in constructing and consistently solving the gap equation and comparing scale dependence across different calculation schemes in the numerical analysis of Sec.~\ref{Sec:Numerics}.

\label{Sec:Analytics}
\subsection{Single-field \texorpdfstring{$\phi ^4$}{} theory} 
Let us consider a simple real $\phi ^4$ theory with a discrete $Z_2$ symmetry. This theory admits phase transition that is second-order and therefore cannot be considered for gravitational wave phenomenology. However, this simplicity also makes the differences between the resummation schemes in obtaining predictions for the thermal potential transparent, and will serve as a useful pedagogical warmup. The tree-level potential has the form,
\begin{equation}
    V_0 = \frac{m^2}{2} \phi ^2 + \frac{g^2}{4!} \phi ^4 \ .
\end{equation}
The one-loop correction at zero temperature in addition to the tree-level potential gives the Coleman-Weinberg potential,
\begin{equation}
    V_{\rm CW} = \frac{M^4}{64 \pi ^2} \left( \log \frac{M^2}{\mu^2} -\frac{3}{2} \right) \ ,
\end{equation}
where $\mu$ is the renormalization scale in the $\overline{\rm MS}$ scheme and $M^2 = m^2 + g^2 \phi ^2/2$ is the field-dependent mass. No physical quantity will depend on the choice of the renormalization scale. Therefore, the above dependence is merely an artefact of the truncation of perturbation theory at finite order. As such, a convenient measure for the convergence of perturbation theory is the residual scale dependence - the implicit scale dependence arising from running couplings in the tree-level potential should cancel the explicit scale dependence in the Coleman-Weinberg potential leaving corrections from the implicit scale dependence in the latter term which is formally higher order. Let us demonstrate this explicitly. Our renormalization group equations at one loop have the form
\begin{eqnarray}
\mu \frac{dm^2}{d\mu} &=& \frac{g^2m^2}{16 \pi ^2} \\
\mu \frac{dg^2}{d\mu} &=& \frac{3g^4}{16 \pi ^2} \ .
\end{eqnarray}
The scale dependence of our potential to $O(g^4/\pi^2)$, dropping two-loop-size terms, then has the form
\begin{eqnarray}
 \mu \frac{dV_0}{d \mu} &=& \frac{g^2 m^2}{32 \pi ^2 } \phi ^2 + \frac{g^4}{128 \pi ^2} \phi ^4 \\
 \mu \frac{dV_{\rm CW}}{d\mu} &=& -\frac{M^4}{32 \pi ^2}
\end{eqnarray}
which sums to a field independent cosmological constant that we can ignore.

At finite temperature, the one-loop correction in the high-temperature expansion has the form,
\begin{equation}
    V_{\rm T} = \frac{1}{48} g^2 T^2 \phi ^2 - T \frac{M^3}{12 \pi} - \frac{M^4}{64 \pi ^2} \log \frac{M^2}{a_b T^2} \ .
\end{equation}
where $a_b=16 \pi ^2 \exp \left(\frac{3}{2}-2 \gamma_{\rm E} \right)$, $\gamma _E$ being the Euler-Mascheroni constant.
It is trivial to see that nothing in our one-loop theory cancels the implicit scale dependence of the finite-temperature piece. Moreover, the size of the uncanceled corrections are as large as the scale dependence of the tree-level pieces in powers of $g/\pi$ and could be relatively larger when $T/\phi$ is large, that is the infrared limit. On top of the explicit infrared divergence that appears at higher-loop, finite-temperature perturbation theory converges slowly in part because of a mismatch of the order of the loop expansion and the size of a term in powers of $g^n/\pi ^m$ as well as the infrared enhancement of uncanceled pieces in terms of $T/\phi$.

\subsubsection{Parwani resummation}
Historically, resumming daisy diagrams was to cure the infrared divergences inherent in finite-temperature field theory \cite{Parwani:1991gq}. However, since this means including higher-loop diagrams in the effective potential, we will also see an improvement in the scale dependence. Parwani resummation  works through replacing $M^2 \to M^2 + \Pi$ where $\Pi = \frac{1}{24} g^2 T^2 $ is the thermal mass to lowest order in the high-temperature expansion. A partial cancellation of the scale dependence occurs
\begin{eqnarray}
 \mu \frac{d V_{\rm T}}{d \mu} &\supset& 3 \frac{g^4}{48 \times 16 \pi ^2}  T^2 \phi ^2 \\ 
 \mu \frac{dV_{\rm CW}}{d \mu} &\supset& - \frac{M^2 \Pi }{16 \pi ^2} \supset -\frac{g^4}{48\times 16 \pi ^2} T^2 \phi ^2 \ .
\end{eqnarray}
The opposite signs in the above terms is the origin of the cancellation. To achieve a full cancellation, we require the missing two-loop term which is the sunset diagram. At high temperature it is
\begin{eqnarray}
 V_{\rm sun} &=& - \frac{1}{12} g^4 \phi ^2 \frac{3T^2}{32 \pi ^4} \left( \log \frac{\mu ^2}{M^2} +2 \right) \nonumber \\
 && \times \left( \frac{\pi ^2}{6} - \frac{\pi M}{2 T} \right) \ .
\end{eqnarray}
In the above we have performed a high-temperature expansion. The full expression that we use in our numerical calculation we put into Appendix \ref{Sunset}. 
A straightforward calculation shows that the scale dependence of the $ T^2 \phi ^2$ term cancels.  The leading order uncanceled piece is
\begin{eqnarray}
 \mu \frac{dV_{\rm Parwani}}{d\mu } = - \frac{g^2T}{ 192 \pi ^2} \left( M^2 + \Pi\right)^{3/2} \ .
\end{eqnarray}
\subsubsection{Dimensional reduction}
At finite temperature in the imaginary time formalism, a thermal field theory is equivalent to a Kaluza-Klein theory with a compactified imaginary time direction of size $1/T$. Integrating out this tower of Matsubara modes leaves one with an effective theory in three dimensions. Doing so automatically includes resummation by construction via the matching relations. Although famously rigorous and consistent, it can be quite formidable technically in a realistic model. In the case of $\phi ^4$ theory, the resulting effective potential is simple enough to be written in a closed form, even at NLO,
\begin{eqnarray}
 V_{\rm DR} &=& T \left[ \frac{1}{2} m_3^2 \phi _3 ^2 + \frac{1}{24} g_3 ^2 \phi _3 ^4 - \frac{1}{12 \pi} \left( M_3^2 \right)^{3/2}  +\frac{1}{16 \pi^2} \right. \nonumber \\ 
&& \left. \times \frac{1}{8} g_3 ^2 M_3^2 + \frac{1}{24} g_3^4 \phi _3 ^2  \frac{1}{16 \pi ^2} \left( 1+ 2 \log  \frac{\mu _3}{3 M_3 } \right)  \right]
\end{eqnarray}
where
\begin{eqnarray}
M_3^2 &=& m_3^2 +\frac{1}{2} g_3^2 v_3^2 \\
g_3^2 &=& T\left(g^2 - \frac{3}{32 \pi ^2} g^4 L_b\right) \\
m_3 ^2  &=& m^2 + \frac{1}{24} g^2T^2 - \frac{1}{16 \pi ^2}\left[ \frac{1}{2} g^2 m^2 L_b +\frac{1}{16} g^2 T^2 L_b \right. \nonumber \\
&& \left. +\frac{1}{6} g_3^4 \left( c+\log \frac{3T}{\mu_3} \right) \right] \\ 
v_3 &=& \frac{\phi}{\sqrt{T}} \ . 
\end{eqnarray}
Here, $\mu _3$ is the scale dependence in the effective theory and $L_b = \left( \log [\mu ^2/T^2] + 2 \gamma _E-2\log 4\pi \right)$. For a fair comparison, we take the larger of the dependencies on $\mu _3$ and $\mu$ as reflective of the residual scale dependence at NLO. To calculate the dependence on the former, it is useful to write
\begin{eqnarray}
 \mu _3 \frac{dm_3^2}{d \mu _3} &=& \mu _3 \frac{dM_3^2}{d \mu _3} \frac{1}{96 \pi ^2} g^4 T^2 \\ 
 \mu _3 \frac{dV_{\rm DR}}{d \mu _3} &\supset &- \frac{1}{384 \pi ^2 }g^4 T^2 \phi ^2 .
\end{eqnarray}
The dependence on $\mu$ turns out to be subdominant and has the form
\begin{eqnarray}
    \mu \frac{d g_3^2}{d \mu}&=& O(g^6) \\ 
    \mu \frac{d m_3^2}{d \mu} &=& \frac{m^2 g^2}{16 \pi ^2 } + \frac{1}{24} \frac{3}{16 \pi ^2} g^4 T^2 \nonumber \\
    && - \frac{1}{16\pi ^2} \left( \frac{1}{8 \pi ^2} g^4 m^2 L_b + g^2 m^2 + \frac{1}{8} g^4 T^2 \right) \\ 
    \mu \frac{d m_3^2}{d \mu} &=& - \frac{1}{16 \pi ^2} \left( \frac{1}{8 \pi ^2} g^4 m^2 L_b  \right) \ ,
\end{eqnarray}
so we focus on the $\mu_3$ dependence and find 
\begin{equation}
    \mu _3 \frac{d V_{\rm DR}}{d \mu _3} = -\frac{1}{768 \pi ^3} g^4 T^3 M_3 \ . 
\end{equation}
This is of order $g^4/\pi ^3$, much smaller than the $\mathcal{O}(g^2)$ residual dependence of the Parwani calculation.

\subsubsection{Gap resummation}
Higher-order diagrams can be included in a simple gap equation where the thermal mass is defined as the second derivative of the resummed potential
\begin{eqnarray}
M^2 &=& m^2 + \frac{g^2}{2}\phi ^2 + \frac{g^2T^2}{24} - \frac{g^2 TM}{8 \pi} \nonumber \\
&& - \frac{g^2 M^2 L_b}{32 \pi ^2 }- \frac{g^4 \phi ^2 L_b}{32 \pi ^2} \ .
\end{eqnarray}
The resulting thermal mass cannot be substituted into the full potential without double counting diagrams. Instead, one includes the resummation of the tadpole ,
\begin{eqnarray}
    V^\prime_{\rm OPD} &=& g^2 \phi \left( \frac{T^2}{24} - \frac{T M}{8 \pi} - \frac{M^2 L_b}{32 \pi ^2} \right) \nonumber \\
    && - \frac{g^4 \phi T}{64 \pi ^2} \left( \log \left[ \frac{\mu ^2}{M^2} \right]+2 \right) \left( \frac{\pi ^2}{6} - \frac{\pi}{2} \frac{M}{T} \right)
\end{eqnarray}
and then integrates over the field to acquire the potential at the end. In the above, everything was written in the high-temperature limit to make the problem analytically tractable. However, the power of gap resummation is that the gap equation can be solved numerically  without a high-temperature expansion.
The scale dependence of the resummed mass and the tadpole can be written to $O(g^4)$,
\begin{eqnarray}
\mu \frac{d V^\prime_{\rm OPD} }{d \mu }&=& \frac{3}{16 \pi ^2 } g^4 \phi \left( - \frac{T M}{8 \pi } \right) + g^2 \phi \left( - \frac{T}{16 \pi M} \mu \frac{d M^2}{d \mu} \right. \nonumber \\
&& \left. - 2 \frac{M^2}{32 \pi ^2} \right) + \frac{1}{384 \pi ^2} g^4 \phi T^2 - \frac{g^4 M T \phi}{64 \pi ^3} \nonumber \\ 
\mu \frac{d M^2}{d \mu ^2} &=& -\frac{g^4 T M}{64 \pi ^3}-\frac{g^2 T}{8 \pi M}\mu \frac{d M^2}{d \mu } +\frac{g^4}{128 \pi ^2}T^2 \ .
\end{eqnarray}
After substituting the derivative of the gap equation with respect to the scale into the scale dependence of the tadpole, the remaining scale dependence cancels up to $O(g^4)$ and the residual piece is $O(g^6)$,
\begin{equation}
    \mu \frac{d V ^\prime_{\rm OPD} }{d \mu } = -\frac{(1-\zeta)g^6 T^3 \phi}{192 \pi ^3 M} + \frac{g^6 T^2 \phi ^2}{256 \pi ^3 M}
\end{equation}
where $\zeta =0$ corresponds to the direct result of the above treatment, and $\zeta=1$ is obtained by including sunsets in the gap equation.
Of course, this is the scale dependence of the tadpole, not the potential. After integration, the residual term is the same order as in DR with a slightly different prefactor, suggesting that the use of gap equations is competitive with NLO DR,
\begin{equation}
    \mu \frac{dV_{\rm OPD}}{d\mu}=\frac{g^4 M T^2(8T(1-\zeta)+3 \phi )}{768 \sqrt{3} \pi^3}-\frac{g^4 m^2 T^2 \phi }{256\sqrt{3}\pi ^3 M} ,
\end{equation} 
where we have, in the final term, expanded an arctanh function that is actually well-behaved in the infrared limit.

\subsection{Two-field \texorpdfstring{$\phi^4$}{} theory}
Let us now consider a two-scalar field model. In principle, this is the minimal model that could produce an observable gravitational wave signature, as the portal couplings can produce a modest thermal barrier. In practice, the peak amplitude tends to be very small. Nevertheless, the predictions of this model can be treated as realistic phenomenological predictions, perhaps existing in some dark sector, under the proviso that any couplings keeping the system in kinetic equilibrium with the visible sector can be sufficiently small that their effect on the potential is negligible. The potential for our model is
\begin{equation}\label{eq:2tree}
    V_0 = \frac{1}{2} m_1^2 \phi _1 ^2 + \frac{g _1^2}{4!} \phi _1 ^4 + \frac{1}{2}m_2^2 \phi _2 ^2 +\frac{g_2^2}{4!} \phi _2 ^4 + \frac{ g_{12}^2}{4} \phi _1 ^2 \phi _2 ^2 \ . 
\end{equation}
We will consider the case where the second scalar does not acquire a VEV throughout the transition, as its function is to provide the thermal barrier. In this case the field dependent masses have the simple form 
\begin{eqnarray}
M_1 ^2 &=& m_1^2 + \frac{1}{2} g_1^2 \phi _1^2 \\ 
M_2 ^2 &=& m_2^2 + \frac{1}{2} g_{12}^2 \phi_1^2 \ .
\end{eqnarray}
Note that we include $\phi_2$ when performing derivatives with respect to the potential and only set $\phi _2$ to zero at the end. To keep equations compact, we do not show their contribution here. 
Finally, the renormalization group equations have the form
\begin{eqnarray}
\mu \frac{d g_1^2}{d \mu } &=& \frac{3 g_1^4}{16 \pi ^2}+ \frac{3 g_{12}^4}{16 \pi ^2} \\ 
\mu \frac{d g_2^2}{d \mu } &=& \frac{3 g_{12}^4}{16 \pi ^2} + \frac{3 g_2^4}{16 \pi ^2} \\
\mu \frac{d g_{12}^2}{d \mu } &=& \frac{g_1^2 g_{12}^2}{16 \pi ^2} + \frac{4 g_{12 }^4}{16 \pi ^2} + \frac{g_{12}^2 g_2^2}{16 \pi ^2} \\ 
\mu \frac{d m_1^2}{d \mu } &=& \frac{g_1^2 m_1^2}{16 \pi ^2} + \frac{g_{12}^2 m_2^2}{16 \pi ^2} \\ 
\mu \frac{d m_2^2}{d \mu} &=& \frac{g_{12}^2 m_1^2}{16 \pi ^2} + \frac{g_2^2 m_2 ^2}{16 \pi ^2}
\end{eqnarray}

\subsubsection{Parwani resummation}
In the Parwani scheme, the thermal masses have the form
\begin{eqnarray}
\Pi _1 &=& \frac{g_1^2 T^2}{24} + \frac{g_{12}^2 T^2}{24} \\
\Pi _2 &=& \frac{g_{12}^2 T^2}{24} + \frac{g_2^2 T^2}{24} \ .
\end{eqnarray}
Let us now put together the scale dependence of each piece of the potential in the Parwani regime. First the tree-level potential,
\begin{equation}
    \mu \frac{d V_0}{d \mu} = \frac{1}{128 \pi ^2} \left( 4 \left( g_1^2 m_1^2 + g_{12}^2 m_2^2 \right) \phi _1^2 + \left( g_1^4 + g_{12}^4 \right) \phi _1^4 \right) \ .
\end{equation}
and the zero-temperature piece of the Coleman Weinberg has the form
\begin{equation}
    \mu \frac{d V_{\rm CW}}{d \mu} \supset -\frac{1}{128 \pi ^2} \left( \left( 2m_1^2 +g_1^2 \phi _1^2  \right)^2 +\left( 2m_2^2 +g_{12}^2 \phi _1^2 \right)^2 \right) \ .
\end{equation}
The above piece cancels the field dependent part of the scale dependence of the tree-level potential.

The quadratic-temperature dependent piece has three parts. First from the high-temperature expansion of the one-loop potential, which at lowest order in the high-temperature expansion is
\begin{equation}
    \mu \frac{d V_{\rm T}}{d \mu} \supset \frac{T^2}{768 \pi ^2}\left(3 g_1^4+7 g_{12}^4+ g_1^2g_{12}^2+g_2^2 g_{12}^2\right) \phi_1^2\ .
\end{equation}
This partially cancels the piece arising from the thermal masses in the Coleman Weinberg potential
\begin{equation}
    \mu \frac{d V_{\rm CW}}{d \mu} \supset -\frac{T^2}{768 \pi ^2}\left( g_1^4 + g_1^2 g_{12}^2 +g_{12}^4+g_{12}^2g_2^2 \right) \phi _1^2 \ . 
\end{equation}
Adding the two terms together leads to a residual piece
\begin{widetext}
\begin{eqnarray}
    \label{e.TFDTVCW}
    \mu \frac{d (V_{\rm T}+V_{\rm CW})}{d \mu } \supset \frac{g_1^4 +3 g_{12}^4}{384 \pi ^2 }T^2 \phi _1^2 -\frac{g_1^2 m_1^2 \bar{M}_1 T+ g_{12}^2 m_1^2 \bar{M}_2 T+g_{12}^2  m_2^2 \bar{M}_1T+g_2^2 m_2^2 \bar{M}_2 T}{128\pi^3} . 
\end{eqnarray}
\end{widetext}
with $\bar{M_i}^2=M_i^2+\Pi_i$, and we also show the next-to-leading $\mathcal{O}(T)$ part of the high-temperature expansion for reasons that will become clear. 
The final contribution is from the leading power sunset term, which in the high-temperature expansion has the form
\begin{eqnarray}
    V_{\rm sun} &=& - \frac{3T^2}{32 \pi ^4}\frac{g_1^4}{12} \phi _1^2 \left( \log \frac{\mu ^2}{ \bar{M_1}^2} + 2 \right) \left( \frac{\pi ^2}{6} - \frac{\pi }{2} \frac{\bar{M_1}}{T} \right) \nonumber \\
    && - \frac{2 T^2}{32 \pi ^4} \frac{g_{12}^4}{4}\phi _1^2 \left( \log \frac{\mu ^2}{\bar{M_2}^2} +2 \right) \left( \frac{\pi^2}{6} - \frac{\pi}{2} \frac{\bar{M_2}}{T} \right) \nonumber \\
    && - \frac{ T^2}{32 \pi ^4} \frac{g_{12}^4}{4}\phi _1^2 \left( \log \frac{\mu ^2}{\bar{M_1}^2} +2 \right) \nonumber \\
    && \times \left( \frac{\pi^2}{6} - \frac{\pi}{2} \frac{\bar{M_1}}{T} \right) \ .
\end{eqnarray}
 The sunset actually cancels the $\mathcal{O}(T^2)$ term in Eq.~({\ref{e.TFDTVCW}}) completely, so we focus on the $\mathcal{O}(T)$ contribution
\begin{equation}
    \label{e.TFDsunset}
    \mu \frac{d V_{\rm sun}}{d \mu} \supset \frac{g_1^4 \bar{M_1} + g_{12}^4(\bar{M_1}+2\bar{M_2})}{128 \pi ^3} T \phi _1^2 \ .
\end{equation}
The residual scale dependence in $V_\mathrm{Parwani}$ is the sum of the $\mathcal{O}(T)$ terms in  Eqs.~(\ref{e.TFDTVCW}) and (\ref{e.TFDsunset}).
If there is strong first-order phase transition, we expect $g_{12}\gg g_1,g_2$. We therefore show the leading order field-dependent term in powers of $g_{12}$:
\begin{equation}
    \mu \frac{d V_{\rm Parwani}}{d \mu} = - \frac{g_{12}^2 (m_2^2 \bar{M_1} + m_1^2 \bar{M_2} )T}{128 \pi ^3} \ .
\end{equation}
The leading order uncanceled term is therefore of the same order as in the single-field case.

\subsubsection{Dimensional reduction}
We use DRalgo \cite{Ekstedt:2022bff} to derive the potential for this model at $O(g^4)$. That is we include two-loop calculations in the dimensionally reduced theory and a NLO (two-loop) resummation (NNLO in the nomenclature of DRalgo). For our analytic comparison, it is easiest to work with the soft rather than the ultrasoft potential, though the appropriate potential to use depends upon where one is in the parameter space. The full soft potential is
\begin{widetext}
\begin{eqnarray}
    V_{3d} &=& \frac{m_{1,3d}^2 \phi _{3d}^2}{2} + \frac{\lambda _{1,3d} \phi _{3d}^4}{24} - \frac{\left(m_{1,3d}^2+\frac{\lambda _{1,3d}\phi _{3d}^2}{2}\right)^{3/2}}{12\pi} - \frac{\left( m_{2,3d}^2 + \frac{\lambda _{12,3d} \phi _{3d}^2}{2}\right)^{3/2}}{12 \pi} +\frac{\lambda _{1,3d} \left( m_{1,3d}^2 + \frac{\lambda _{1,3d} \phi _{3d}^2}{2} \right) }{128 \pi ^2} \nonumber \\
    &&  + \frac{\lambda_{12,3d} \sqrt{m_{1,3s}^2 + \frac{\lambda _{1,3d} \phi_{3d}^2}{2}} \sqrt{m_{2,3s}^2 + \frac{\lambda _{12,3d} \phi_{3d}^2}{2}} }{64 \pi ^2} +\frac{\lambda _{2,3d} \left(m_{2,3s}^2 + \frac{\lambda _{12,3d} \phi_{3d}^2}{2} \right)}{128 \pi ^2} \nonumber \\ 
    && - \frac{\lambda _{1,3d}^2 \phi _{3d}^2 \left( \frac{1}{2} + \log \left[ \frac{\mu _3}{3\sqrt{m_{1,3s}^2 + \frac{\lambda _{1,3d} \phi_{3d}^2}{2}}}\right] \right)}{192 \pi ^2} +\frac{\lambda _{12,3d} ^2 \phi _{3d} ^2 \left( \frac{1}{2} + \log \left[ \frac{\mu _3}{\sqrt{m_{1,3s}^2 + \frac{\lambda _{1,3d} \phi_{3d}^2}{2}} +2 m_{2,3s}^2 + \frac{\lambda _{12,3d} \phi_{3d}^2}{2}}\right] \right) }{64 \pi ^2} \label{eq:nlodr} \ .
\end{eqnarray}
\end{widetext}
Here the dimensionally reduced couplings, masses and fields are given by
\begin{eqnarray}
    \lambda_{1,3d} &=& T \left( g_1^2 - \frac{3 L_b (g_1^4 + g_{12}^4)}{32 \pi ^2} \right) \\
    \lambda _{2, 3d} &=& T \left(g_2^2 - \frac{3 L_b (g_{12}^4+g_2^4)}{32 \pi ^2} \right) \\
    \lambda_{12,3d} &=& T \left( g_{12}^2 - \frac{L_b g_{12}^2 (g_1^2 + 4 g_{12}^2 +g_2)^2}{32 \pi ^2}\right) 
\end{eqnarray}
\begin{eqnarray}
    m_{1,3d}^2 &=& m_1^2 + \frac{1}{24}T^2 (g_1^2 + g_{12}^2) \nonumber \\ 
    && -\frac{1}{768 \pi ^2} \left( L_b \left[ 24 m_1^2 g_1^2 +24 m_2 ^2 g_{12}^2 \right. \right. \nonumber \\
    && \left. \left. + T^2 (-g_1^4 + g_1^2 g_{12}^2 + g_{12}^2 [-5 g_{12}^2 + g_2^2]) \right] \right. \nonumber \\ 
    && \left. +4 T^2 (g_1^4+ 3g _{12}^4)(\gamma _E - 12 \log A ) \right. \nonumber \\
    && \left. - 8 (3 \lambda _{12,3d}^2 + \lambda_{1,3d}^2) \log \frac{\mu _3}{\mu} \right)
\end{eqnarray}
\begin{eqnarray}
    m_{2,3d}^2 &=& m_2^2 + \frac{1}{24}T^2 (g_2^2 + g_{12}^2) \nonumber \\ 
    && -\frac{1}{768 \pi ^2} \left( L_b \left[ 24 m_1^2 g_{12}^2 +24 m_2 ^2 g_{2}^2 \right. \right. \nonumber \\
    && \left. \left. + T^2 (-g_2^4 + (g_1^2+g_2^2) g_{12}^2 -5 g_{12}^4) \right] \right. \nonumber \\ 
    && \left. +4 T^2 (g_2^4+ 3g _{12}^4)(\gamma _E - 12 \log A ) \right. \nonumber \\
    && \left. - 8 (3 \lambda _{12,3d}^2 + \lambda_{2,3d}^2) \log \frac{\mu _3}{\mu} \right) \\ 
    \phi _{3d} &=& \phi _1 / \sqrt{T},
\end{eqnarray}
respectively. Again, $\mu _3$ is the renormalization scale in the three-dimensional theory and $A\sim 1.2$ is the Glaisher number. It is straight forward to show that the residual scale dependence of $(\lambda _{1,3d},\lambda _{2,3d},\lambda _{12,3d})$ is of $O(g^6_i)$. The masses cancel at the same order with the exceptions
\begin{eqnarray}
 \mu _3    \frac{\partial m_{1,3d}^2}{\partial \mu _3} &=& \frac{(g_1^4+3 g_{12}^4)T^2}{96 \pi ^2} \\
  \mu  _3  \frac{\partial m_{2,3d}^2}{\partial \mu _3} &=& \frac{(g_2^4+3 g_{12}^4)T^2}{96 \pi ^2} \ .
\end{eqnarray}
The residual scale dependence then for NLO DR is
\begin{equation}
    \label{e.NLODRscaledep}
    \mu _3 \frac{\partial V_{\rm DR}}{\partial \mu _3} = - \frac{(g_1^4 \bar{M_1} + g_2^4 \bar{M_2} + 3 g_{12} ^4 (\bar{M_1} + \bar{M_2} ))T^3}{768 \pi ^3} 
\end{equation}
where
\begin{eqnarray}
    \bar{M_1}^2 &=& m_1 ^2 + \frac{g_1 ^2}{2} \phi ^2 + \frac{g_1^2 T^2}{24} + \frac{g_{12}^2 T^2}{24} \nonumber \\     
    \bar{M_2}^2 &=& m_2 ^2 + \frac{g_2 ^2}{2} \phi ^2+ \frac{g_2^2 T^2}{24} + \frac{g_{12}^2 T^2}{24}
\end{eqnarray}
This again looks to be of similar order as in the single-field case, fourth order in the couplings as opposed to second order for Parwani resummation. 

\subsubsection{Gap resummation}
In this model we have to consider two coupled gap equations, one for each scalar field,

\begin{eqnarray}
    M_1^2 &=& m_1^2+\frac{1}{2}g_1^2\phi_1^2+\frac{g_1^2T^2}{24}+\frac{g_{12}^2T^2}{24}- \frac{g_1^2 M_1 T}{8 \pi} - \frac{g_{12}^2M_2 T}{8 \pi} \nonumber \\ 
    && -\frac{g_1^2 M_1^2 L_b}{32 \pi ^2} - \frac{g_{12}^2 M_2^2 L_b}{32 \pi ^2} -\frac{g_{1}^4 L_b \phi _1^2}{32 \pi ^2} - \frac{g_{12} ^4 L_b \phi _1^2}{32 \pi ^2} \nonumber \\
    && - \frac{g_1^4 T \phi _1^2}{16 \pi M_1} - \frac{g_{12} ^4 T \phi _1^2}{16 \pi M_2 } \\
    M_2 ^2 &=& m_2^2+\frac{1}{2}g_{12}^2 \phi^2+\frac{g_{12}^2T^2}{24}+ \frac{g_2^2 T^2}{24}-\frac{g_{12}^2 M_1 T}{8 \pi} - \frac{g_2^2 M_2 T}{8 \pi} \nonumber \\
    &&-\frac{g_{12}^2 L_b M_1^2 }{32 \pi ^2} - \frac{g_2^2 L_b M_2^2}{32 \pi ^2} \ .
\end{eqnarray}
Excluding the sunset, the one-loop tadpole has the form
\begin{eqnarray}
    V^\prime_{\rm 1} &=& \frac{1}{24 }g_1^2 T^2 \phi_1 + \frac{1}{24} g_{12}^2 T^2 \phi_1 -\frac{g_{1}^2 M_1 T \phi_1}{8 \pi} - \frac{g_{12}^2 M_2 \phi_1}{8 \pi} \nonumber \\
    && - \frac{g_1^2 M_1^2 \phi_1 L_b}{32 \pi ^2} - \frac{g_{12}^2 M_2^2 \phi_1 L_b}{32 \pi ^2}\ .
\end{eqnarray}
Let us first consider the scale dependence of the gap equations up to $O(g^4)$,
\begin{eqnarray}
    \mu \frac{d M_1^2}{d \mu} &&= \frac{g_1^4 T^2+3g_{12}^4 T^2}{192\pi^2} \nonumber \\
    && -\frac{3(g_1^4+g_{12}^4)M_1 T+6 g_{12} ^4M_2 T+12 g_1^4 \phi _1^2}{192\pi^3}\\
    \mu \frac{d M_2^2}{d \mu} &&= \frac{3 g_{12}^4T^2 +g_2^4 T^2}{192 \pi^2} \nonumber \\
    && - \frac{3(g_2^4 +g_{12}^4) M_2 T + 6 g_{12}^4 M_1 T +24 \pi \phi _1^2}{192 \pi ^3} \ .
\end{eqnarray}
The leading terms in the tadpole are
\begin{eqnarray}    
    \mu \frac{d V^\prime_1}{d \mu} &=& \frac{(g_1^4 +3 g_{12}^4)T^2 \phi _1}{192 \pi ^2} \nonumber \\
    && -\frac{(3g_1^4M_1 +3 g_{12}^4 (M_1 + 2 M_2))T\phi _1}{192 \pi ^3} \nonumber \\
    && + \mu \frac{ d V^\prime_{\rm sun}}{d \mu}
\end{eqnarray}

It is straight forward to see that the sunset terms cancels the above exactly. The remaining $O(g_{12}^6)$ term is a little cumbersome, so we omit the explicit expression. However, after integrating, the residual piece is $O (g_{12}^4/\pi ^3)$, which is the same order the uncanceled piece in dimensional reduction,  Eq.~(\ref{e.NLODRscaledep}).

Note again how DR and OPD perform similarly, and both much better than Parwani resummation, with the size of scale dependence parametrically reduced by two powers of the coupling.
Our derivation also makes apparent the usability advantage of OPD. Even for this simple toy theory, the OPD calculation is much more tractable than dimensional reduction, and one does not need to take care about whether to use the soft or ultrasoft potential.

\section{Numerical Implementation of Gap Resummation and Results}
\label{Sec:Numerics}

We now review how to set up the numerical OPD calculation. In particular, we specify that an iterative approach should be used without any mass derivatives in the gap equation, clarifying some ambiguities from the original numerical treatment~\cite{Curtin:2016urg}. An important point to note is that in our numerical implementation we will not rely on high-temperature approximation and will be solving the gap equation away from the origin, neither of which was done in the previous section. We then present numerical results for the thermal potential and its gravitational wave observables for a representative benchmark point in the two-field $\phi^4$ theory, comparing OPD and Parwani resummation to demonstrate the improved accuracy and precision of the OPD calculation. 
To facilitate comparison to dimensional reduction, which only works when the high-temperature approximation is valid, we also compare effective potentials for a scenario that does not yield a strong phase transition, showing that the OPD calculation yields a result very close to the dimensional reduction calculation, both of which differ from the Parwani result. This supports the expectation from our analytical derivation that one-loop OPD is competitive with two-loop dimensional reduction, and motivates development of RG improvement for OPD to fully realize the OPD accuracy promised by the analytical results of the previous section. (As explained in the introduction, these numerical studies all utilize effective potentials without RG improvement, as formulating a consistent RG-improvement scheme for OPD that is not restricted to the high-temperature approximation is the subject of upcoming work.)

\subsection{Construction and solution of gap equation away from origin}

Our procedure for OPD resummation is as follows. Note we restrict ourselves here to the scenario where only one scalar acquires a VEV during the phase transition:
\begin{itemize}
    \item We use $\delta m^2_i$ instead of $\Pi_i$ to denote corrections to each scalar field's mass beyond the tree-level $m_i^2$, since it in general includes both zero- and finite-temperature corrections.  For a given temperature, the mass corrections $\delta m^2_{i}$ are obtained by numerically solving a set of coupled algebraic gap equations on a grid of field values along the excursion of the symmetry breaking field, in this case $\phi_1$,
    \begin{widetext}
    \begin{equation} \label{eq:gap}
        \delta m^2_{j}(\phi_1,T)=\sum_{i} \left[ \frac{\partial ^2 V^{i}_{\text{CW}}}{\partial \phi_j^2} \left( m_{i}^2(\phi_1)+\delta m^2_{i}(\phi_1,T)  \right) + \frac{\partial ^2 V^{i}_{\text{th}}}{\partial \phi_j^2} \left( m_{i}^2(\phi_1)+\delta m^2_{i}(\phi_1,T)  \right) \right]
    \end{equation}
    \end{widetext}
    \item The continuous functions $\delta m^2_{j}(\phi_1,T)$ are obtained by interpolating the solutions to the gap equation on a grid of $\phi_1$ VEVs, which are then substituted in the first derivative of the zero + finite-temperature one-loop potential, plus the finite-temperature two-loop sunset term.
    \begin{widetext}
    \begin{eqnarray}
    V_{\text{OPD}} &=& V_0 + \sum_{i} \int d \phi_1 \left[ \frac{\partial V^{i}_{\text{CW}}}{\partial \phi_1} \left( m_{i}^2(\phi_1)+\delta m^2_{i}(\phi_1,T)  \right) \right. \nonumber \\
    && \left. + \frac{\partial V^{i}_{\text{th}}}{\partial \phi_1} \left( m_{i}^2(\phi_1)+\delta m^2_{i}(\phi_1,T)  \right) +\frac{\partial V^{i}_{\text{sun}}}{\partial \phi_1} \left( m_{i}^2(\phi_1)+\delta m^2_{i}(\phi_1,T)  \right) \right]
    \end{eqnarray}
    \end{widetext}
\end{itemize}
One of the crucial aspects of OPD is its numerical efficiency while going beyond the high-temperature approximation, achieved by using full thermal integrals in the potential but high-temperature approximations in the gap equation.\footnote{Note that the full thermal integrals can be efficiently approximated to high precision by a piecewise defined function joining the high- and low-T approximations, which for the bosonic case is given by $J_B^{\rm piecewise}(y^2)= J_B^{\rm high-T}(y^2)$ $\left[  -\sum \limits_{n=1}^{3}\frac{y^2}{n^2}K_2(y\, n) \right]$ for $y^2$ less [more] than 0.22, where $K_2$ is the modified Bessel function of the second kind. We limit $n \leq 3$ but more precision is easily obtained by including more terms.}
This was justified in~\cite{Curtin:2016urg} by arguing that the gap equation only matters for phase transitions in regions of field- and parameter-space where components of the plasma become light and hence the high-temperature expansion is valid, while further out in field space where the high-temperature expansion fails the thermal mass is accurately treated as small in the full potential, meaning the gap equation has a much smaller effect and its error is parametrically suppressed.
As we outline in the next subsection, we have systematically verified that this assumption is in fact correct, justifying the use of the high-temperature expansion in the gap equation and the enormous simplification it brings.

The algebraic equation (\ref{eq:gap}) can be solved iteratively:
\begin{widetext}
\begin{equation}
\label{e.iterative}
        \delta m^2_{j}(\phi_1,T)_{n+1}=\sum_{i} \left[ \frac{\partial ^2 V^{i}_{\text{CW}}}{\partial \phi_j^2} \left( m_{i}^2(\phi_1)+\delta m^2_{i}(\phi_1,T)_{n}  \right) + \frac{\partial ^2 V^{i}_{\text{th}}}{\partial \phi_j^2} \left( m_{i}^2(\phi_1)+\delta m^2_{i}(\phi_1,T)_{n}  \right) \right]
\end{equation}
\end{widetext}
for a fixed $\phi_1$ and $T$ where the solution starts at $\delta m^2_{j}(\phi_1,T)_{n=1}=0$ and converges to a fixed value after a handful of iterations.

The above is almost identical to the numerical OPD procedure originally laid out in~\cite{Curtin:2016urg}, with one important difference. The authors of ~\cite{Curtin:2016urg} found that the purely algebraic, iterative method of solution is problematic, since it yielded multiple oscillating solutions to the gap equations away from the origin in field space when applied to the SM with simple scalar extensions. This was solved in an \emph{ad hoc} manner, by using an alternative formulation of gap equations which involved keeping mass derivatives in the gap equation and using them to constrain $\delta m_j(\phi_1 + \Delta \phi_1, T)$ based on the previous solution  $\delta m_j(\phi_1, T)$ on the $\phi_1$-grid, turning the algebraic gap equations into a set of differential equations. While this yielded unique and apparently reasonable solutions most of the time, the procedure was very vulnerable to numerical errors due to the singular nature of the resulting gap equation near field values where $\partial^2 V/\partial\phi_1^2$ flips sign, i.e. when passing through the potential barrier at the critical temperature. The inclusion of derivative terms was also not strictly justified by the original proof of the validity of partial dressing in~\cite{Boyd:1993tz}, even though their effect was expected to be subleading.

In our careful analysis of OPD as applied to the much simpler two-field $\phi^4$  theory, we found that this differential version of the gap equation yielded numerical solutions of the effective potential that contained unacceptable artifacts for the sizeable couplings that yield a first-order phase transition. On the other hand, the simpler iterative approach always yielded unique and reasonable solutions to the system of gap equations. We therefore use the solution method of Eq.~(\ref{e.iterative}) in our analysis.
We hypothesize that the nonconvergence of the gap equation solution in~\cite{Curtin:2016urg} was caused by applying OPD to the full SM with extra scalars, without consistently including gauge bosons in the system of gap equations (rather just including their $\mathcal{O}(T^2)$ contributions in the scalar gap equations).

Our purely local and iterative  implementation of the gap equation is therefore fully consistent with the original formulation of partial dressing~\cite{Boyd:1993tz}. This is ultimately a fortunate development for the practical application of OPD, as the purely iterative numerical implmentation is much simpler and faster. We will show how to consistently expand OPD to include gauge bosons in an upcoming publication.

\begin{table}
    \centering
    \begin{tabular}{||c|c|c|c||}
    \hline
    \multicolumn{2}{||c|}{Physical parameters}  & \multicolumn{2}{|c||}{$\overline{\text{MS}}$ parameters}\\
    \hline
      $\langle \phi_1 \rangle$   &  400 GeV & \quad $\mu_1$ \quad & 50.12 GeV \\
      $M_{1,{\rm pole}}$   & 125 GeV & \quad $m_2$ \quad & 193.32 GeV \\
      $M_{2,{\rm pole}}$   & 600 GeV & \quad $g_1^2$ \quad & 0.28 \\
      $g_{12,{\rm phys}}^2$ & 4.40 & \quad $g_{12}^2$ \quad & 4.37 \\
      $g_{2,{\rm phys}}^2$ & 0.60 & \quad $g_{2}^2$ \quad & 1.12 \\
    \hline
    \end{tabular}
    \caption{Physical and $\overline{\text{MS}}$ parameters of the two-field model for our benchmark numerical calculation at the renormalization scale $\mu=600$ GeV (check Appendix \ref{Sec:AppA} for the relation between the two). The fixed value of the renormalization scale is only used to plot Figs.~\ref{fig:pots} and ~\ref{fig:maxomega}.}
    \label{Parameter_table}
\end{table}

\subsection{Numerical results for benchmark two-field \texorpdfstring{$\phi^4$}{} theory}

We will now consider the two-field $\phi^4$ theory with benchmark parameters shown in Table~\ref{Parameter_table}, numerically computing the effective potential and gravitational wave signal in Parwani and OPD resummation to compare the two schemes.  %
We find similar behaviour for other parameter points with strong phase transitions, so these results are representative. 

This numerical study is done by choosing the $\overline{{\rm MS}}$ parameters at zero temperature for an arbitrary renormalization scale $\mu$ (chosen to lie near the $\phi_2$ mass) which reproduces experimental observables at one-loop and using these parameters to compute the thermal potential and gravitational signal. This procedure is then repeated for different choices of $\mu$ to estimate the scale dependence and hence theoretical uncertainty of our results ~\cite{Gould:2021oba}. The experimental observables for our model are the vev $ \langle \phi_1 \rangle$, pole mass $M_1, M_2$ of the scalar fields  at the symmetry broken vacuum and the quartic couplings $g_{12}^2$ and $g_2^2$. The details of the one-loop matching of the parameters to the experimental observables are given in Appendix \ref{Sec:AppA}. 

\begin{figure}
    \centering
    \includegraphics[width=0.45\textwidth]{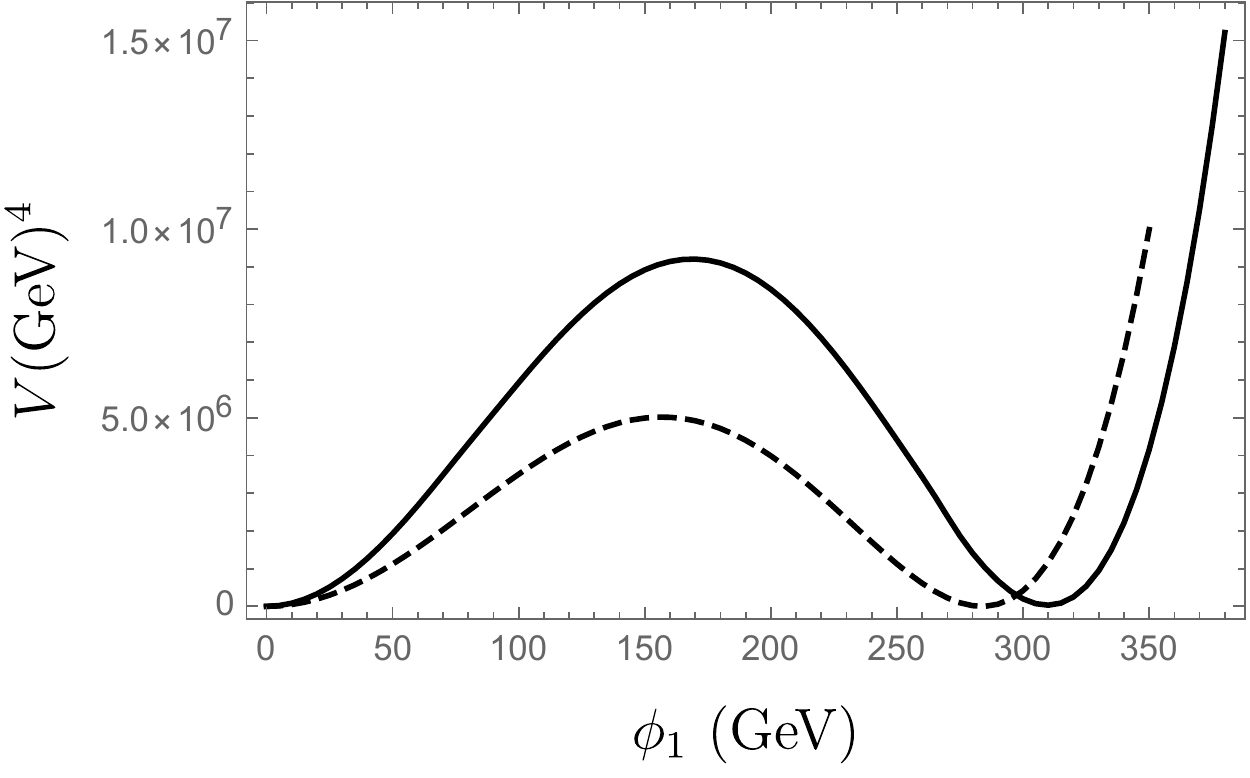}
    \includegraphics[width=0.45\textwidth]{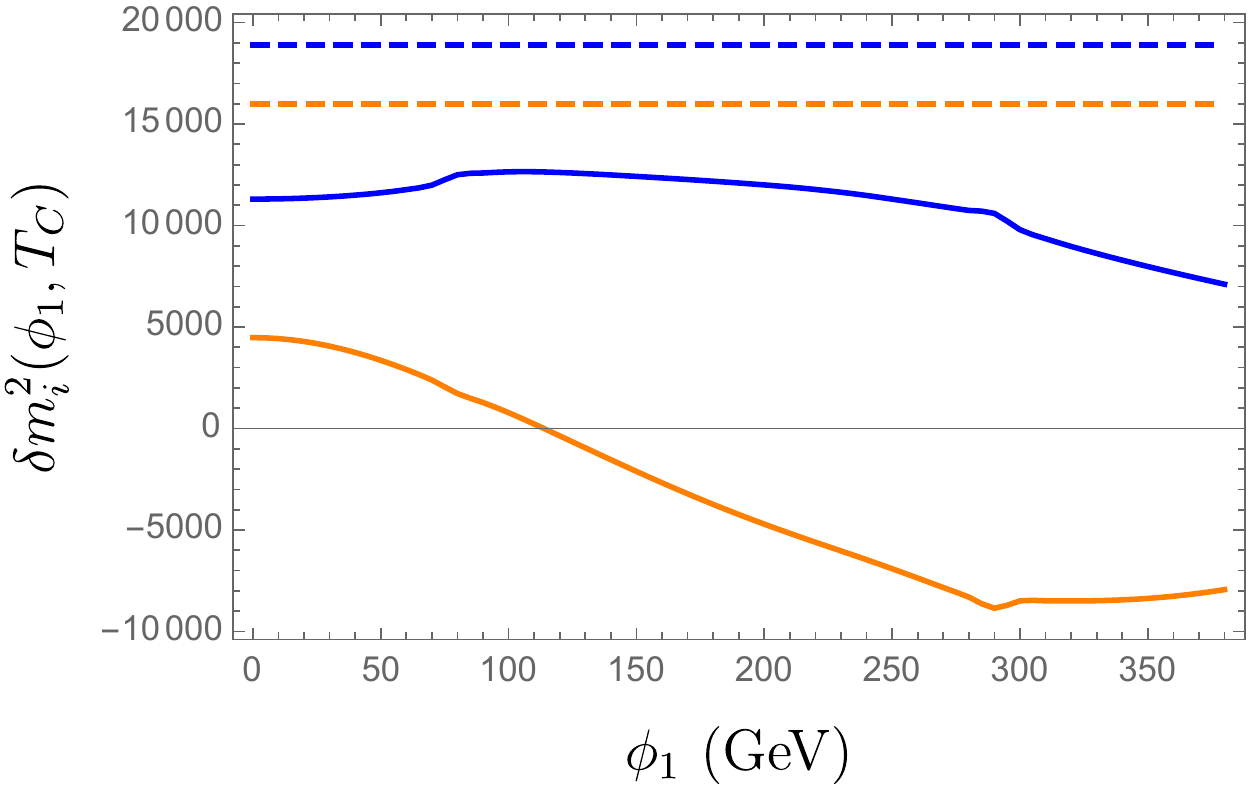}
    \caption{Upper panel: The effective potential with one-loop matching for the benchmark in Table~\ref{Parameter_table} and $\mu = 600$ GeV with the solid line denoting OPD and the dashed line referring to Parwani resummation. Lower panel: Numerical solution
    of the gap equation for the mass correction $\delta m_i^2(\phi_1, T_c)$ at the critical temperature for the same benchmark. Here $T_c=287$ GeV for Parwani and $T_c=265.6$ GeV for OPD. The orange and blue curves are for $\phi _1$ and $\phi _2$ respectively. }
    \label{fig:pots}
\end{figure}

Figure \ref{fig:pots} shows the thermal potential for our benchmark point at $T = T_c$ when the true and false minima are degenerate, as well as the corresponding solutions to the gap equation. As one can see the iterative method does give a smooth, unique solution exhibiting correct physical behavior, whereby the mass corrections are maximum at the origin and decreases with $\langle \phi_1 \rangle$ since the fields acquire more mass reducing their participation in the thermal plasma. This is markedly different from the constant thermal masses assumed in Parwani resummation, which makes use of the lowest-order high-temperature expansion far away from the origin where it is no longer justified.

On the other hand, we also verified, for this and other choices of parameters, that the use of the high-T expansion in the gap equation for OPD was valid. Compared to solving the gap equations with full thermal functions, we only found meaningful differences in the region of field space where $M^2/T^2$ is large, $M^2$ being the resummed mass. For example, for this particular benchmark this difference shows up when $M_2^2/T^2 \gtrsim 3$, significantly larger than the still sizeable but more moderate values relevant for our phase transition. Even then, the difference in $\delta m_i^2$ (full thermal potential) obtained with the high-temperature approximation in the gap equation vs full thermal functions in the gap equations are at most $\sim10\%$ (1\%). This confirms the original argument for the high-temperature approximation in the gap equation made in~\cite{Curtin:2016urg}.

\begin{figure}
    \centering
    \includegraphics[width=0.45\textwidth]{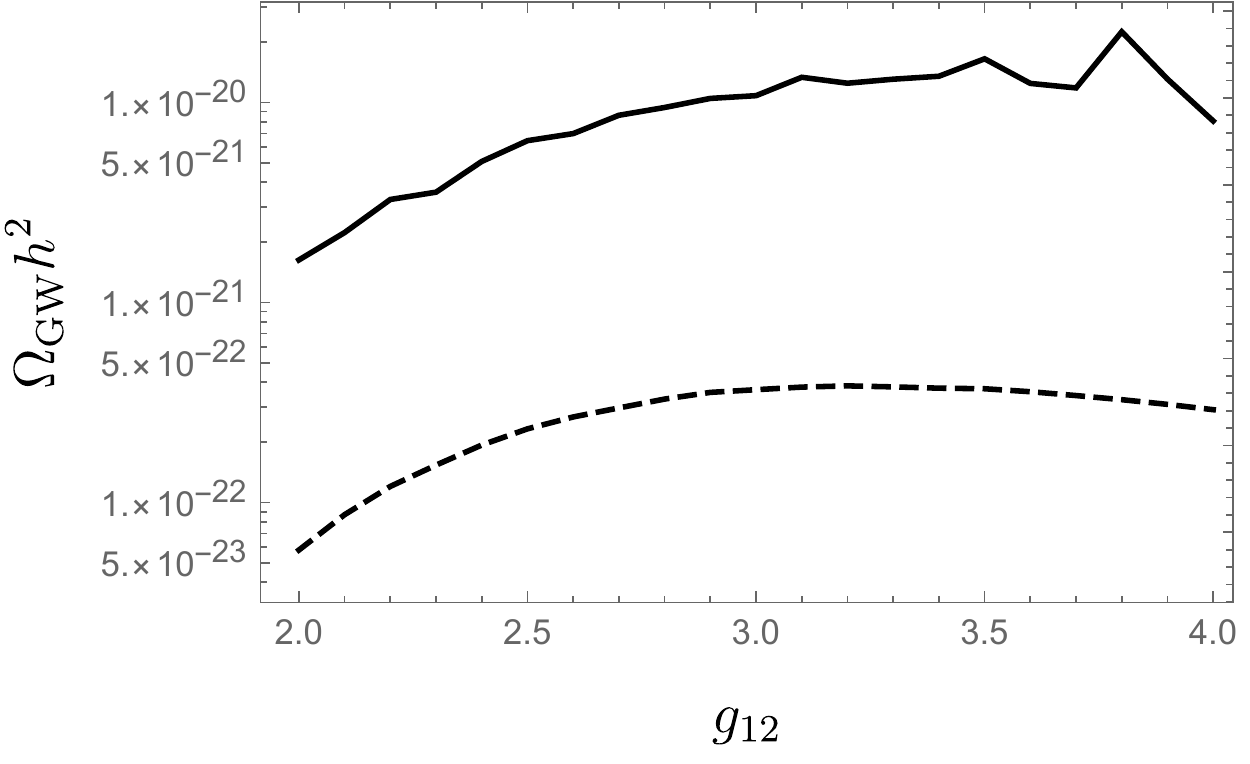}
    \caption{The maximal gravitational wave amplitude for all parameters fixed except $g_{12}$. Solid line denotes the prediction of OPD and dashed line the predictions of Parwani.}
    \label{fig:maxomega}
\end{figure}

OPD by construction is more accurate than Parwani, as it takes into account both proper counting of diagrams and higher-order corrections. These higher-order corrections do make a significant change in the profile of the thermal potential. OPD seems to predict a larger barrier for the thermal potential when compared to Parwani as seen in Fig. \ref{fig:pots}. This tends to be a general feature of OPD throughout the parameter space and results in the maximal gravitational wave amplitude being orders of magnitude larger than what is predicted by Parwani as can be seen in Fig. \ref{fig:maxomega}. 

\begin{figure}[h!]
    \centering
    \includegraphics[width=0.45\textwidth]{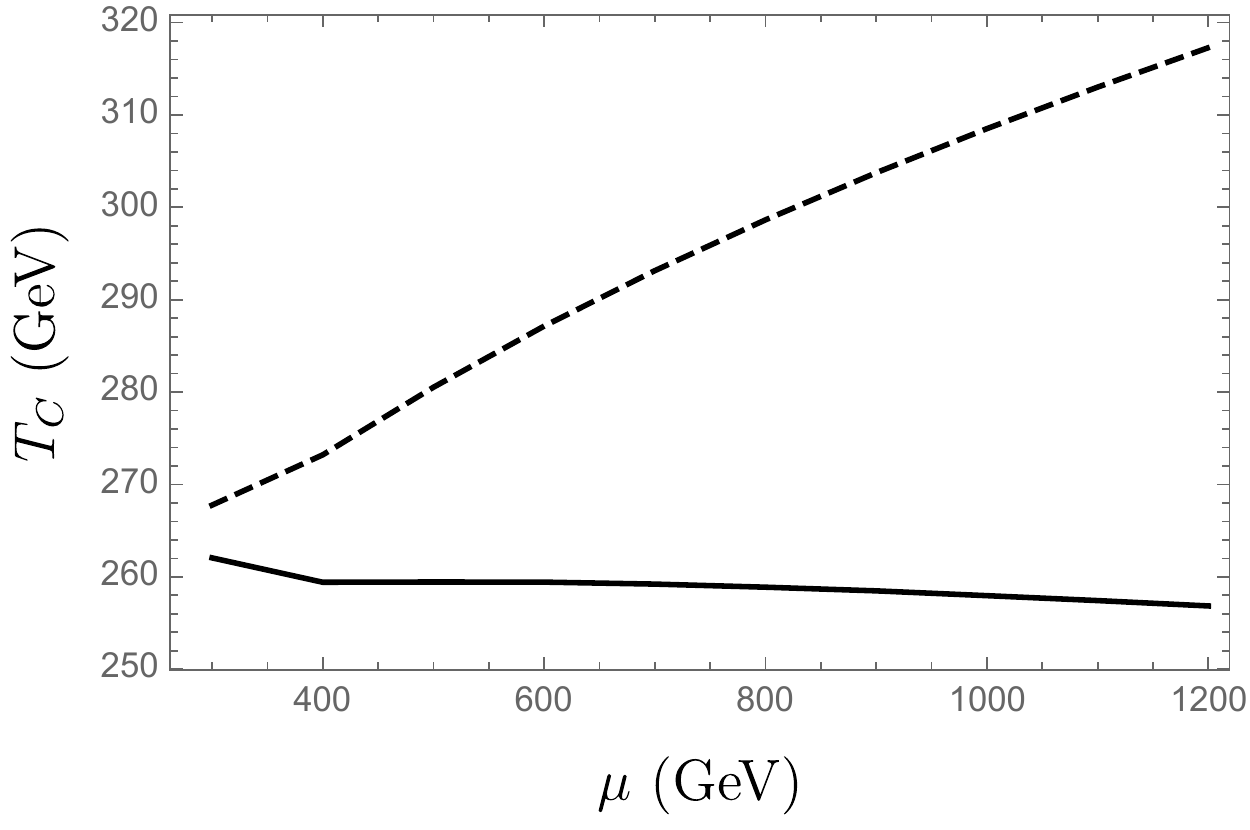}
    \includegraphics[width=0.45\textwidth]{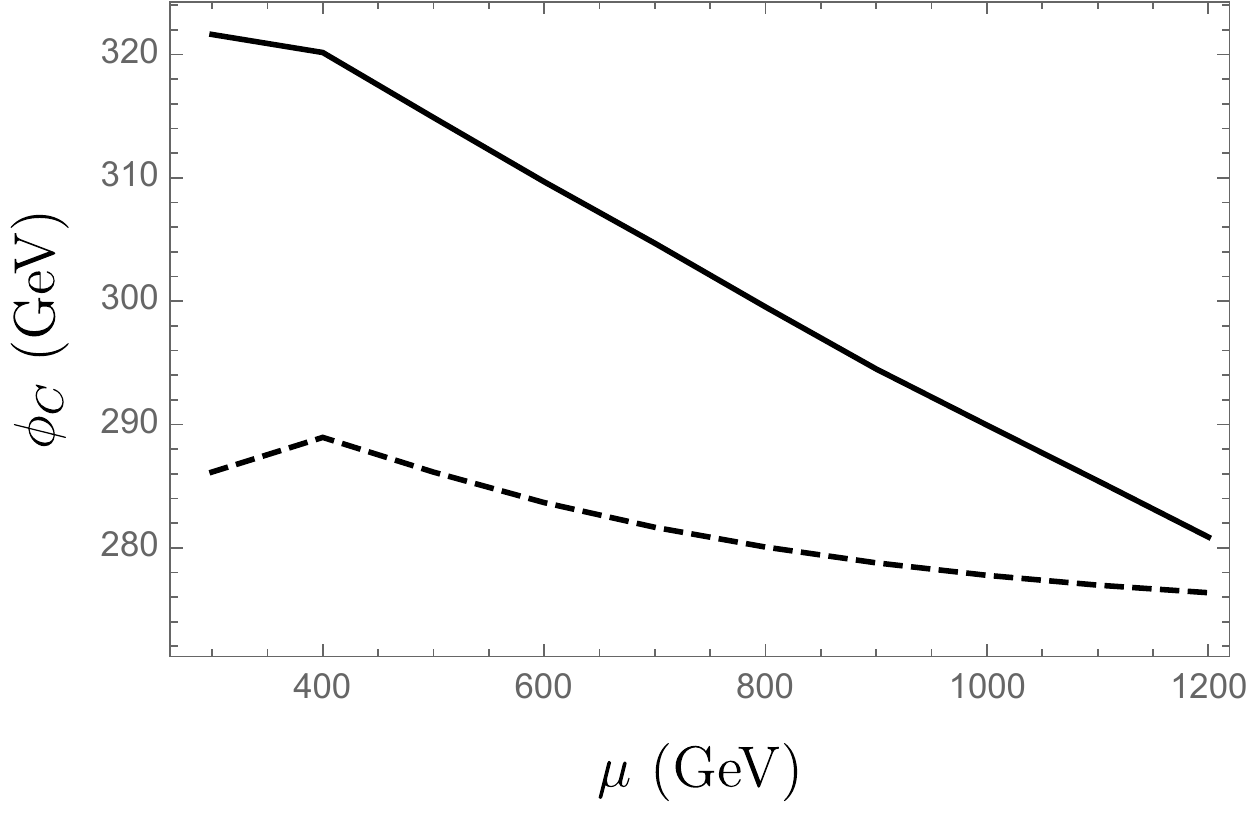}
    \caption{The scale dependence ($\mu$) of the critical temperature and VEV for the benchmark in Table \ref{Parameter_table} with dashed line corresponding to Parwani and solid line corresponding to OPD with both augmented by two-loop sunsets.}
    \label{fig:phictc}
\end{figure}

\begin{figure}[h!]
    \centering
    \includegraphics[width=0.45\textwidth]{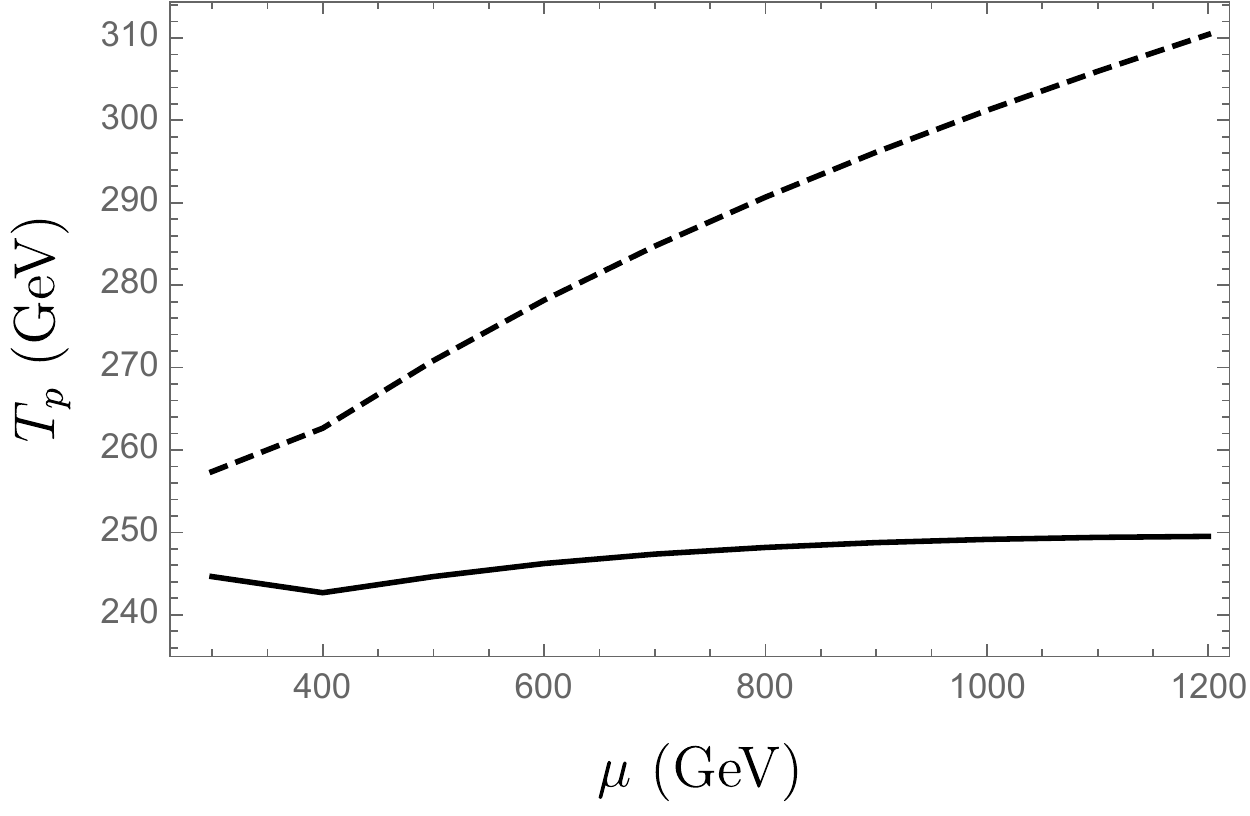}
    \includegraphics[width=0.45\textwidth]{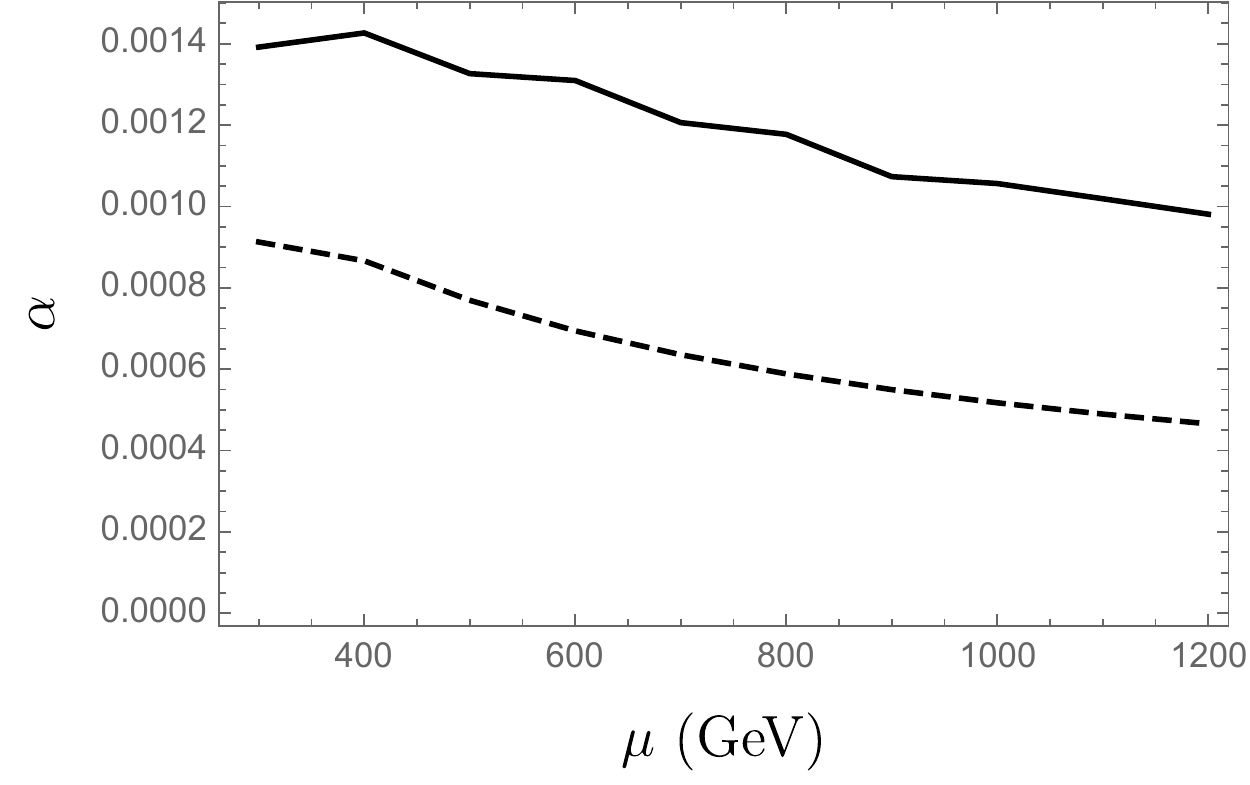} \\ 
    \includegraphics[width=0.45\textwidth]{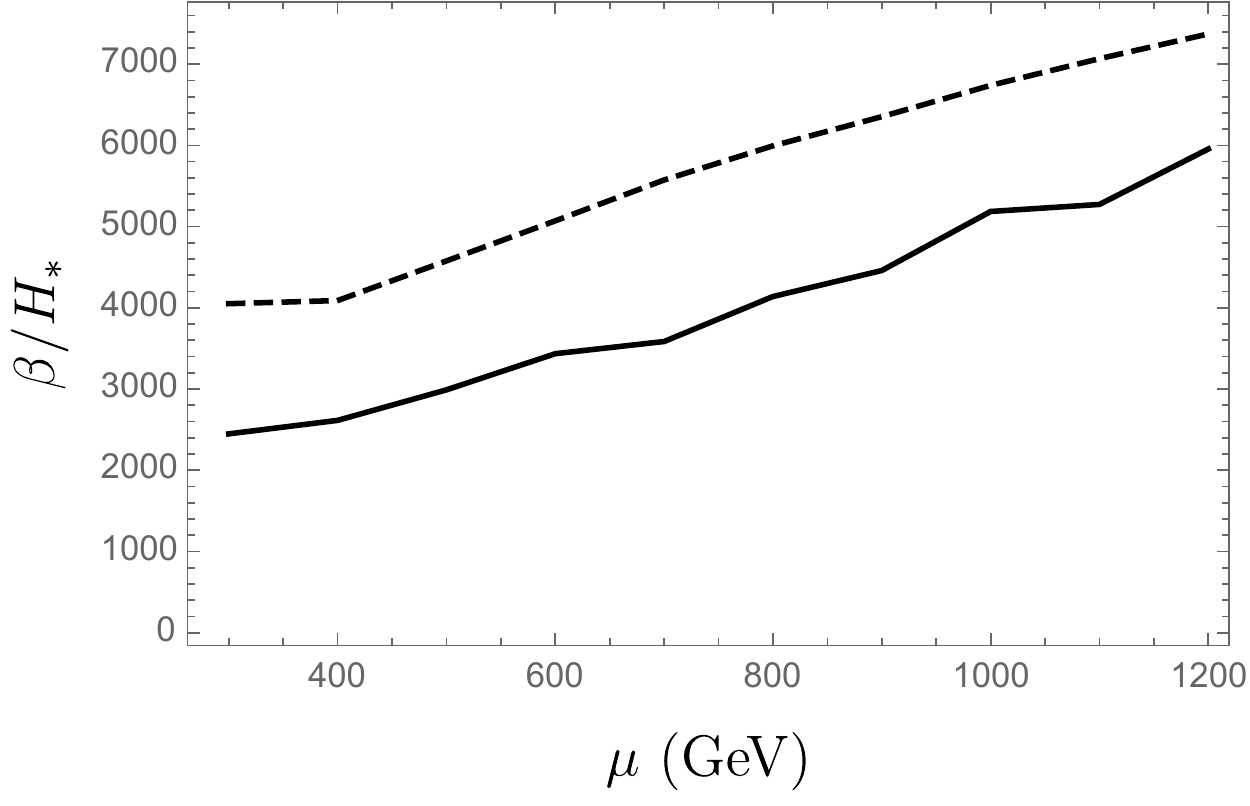}
    \includegraphics[width=0.45\textwidth]{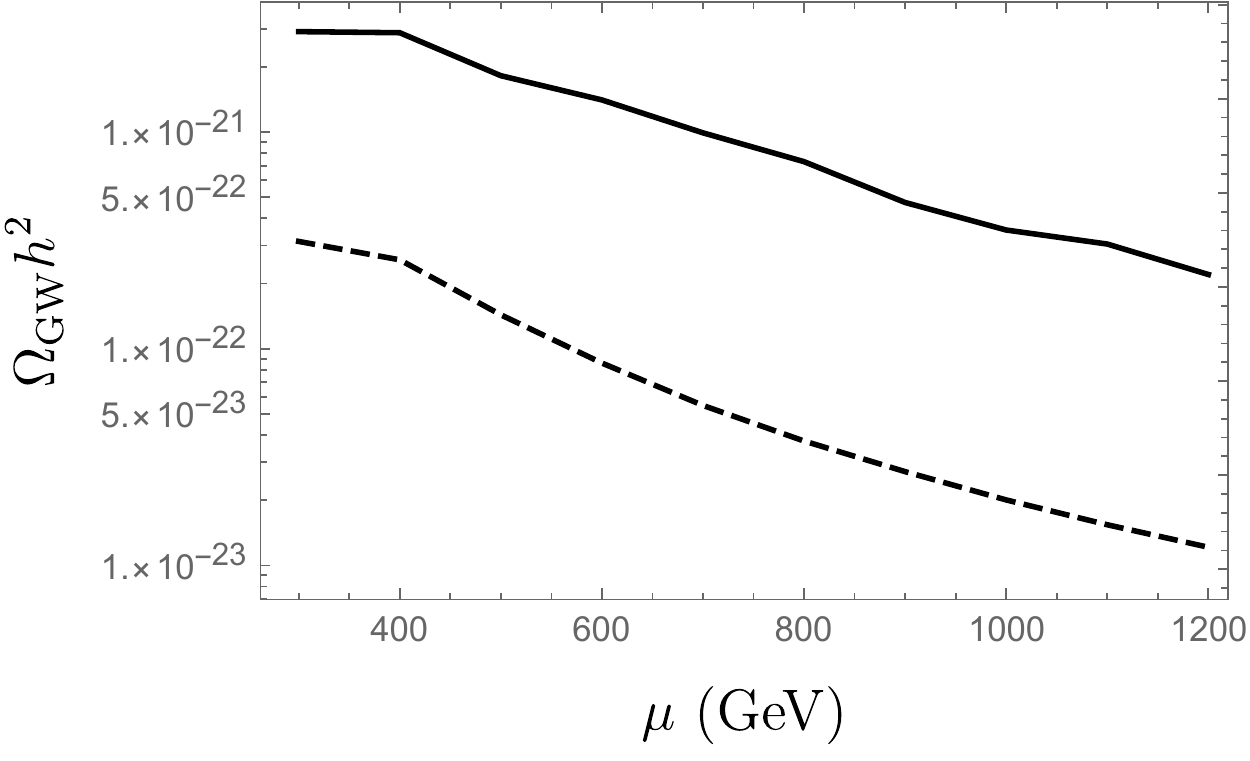}
    \caption{Scale dependence ($\mu$) of the thermal parameters and the peak gravitational wave amplitude as predicted by the sound shell model. Solid/dashed lines correspond to OPD/Parwani, respectively.
    }
    \label{fig:GWs}
\end{figure}

In addition to better accuracy, the numerical results also demonstrates significantly improved precision, as signaled by the reduction of scale dependence when using OPD. Figures \ref{fig:phictc} and \ref{fig:GWs} show the variation of the thermal parameters and peak gravitational wave amplitude as a function of the renormalization scale. An estimate of the variation of the critical and percolation temperature , $\Delta T =\frac{T_{\rm max}-T_{\rm min}}{T_{\rm max}}$ on varying the renormalization scale gives $\Delta T_{\rm OPD} \sim 2 \%$ compared to $\Delta T_{\rm Parwani} \sim 15 \%$. Similarly, almost a factor of 1.7 reduction is observed in the variation of the strength of the phase transition, $\Delta \alpha$ with $\Delta \alpha_{\rm Parwani} \sim 49 \%$ and $\Delta \alpha_{\rm OPD} \sim 29 \%$. Of course, it is to be noted that not all the thermal observables show such an improvement. In particular it can be seen in Fig. \ref{fig:phictc} that $\phi_c$ has a larger variation in OPD ($\Delta \phi_{c,{\rm OPD}} \sim 12.6 \% $) compared to Parwani ($\Delta \phi_{c,{\rm Parwani}} \sim 3.4 \% $). This on the other hand has minimal effect on variation of the gravitational wave observables since both $\alpha$ and $\beta / H_* $ have stronger dependence on $T_p$ rather than $\phi_c$.
In particular, this also means that OPD's prediction of the gravitational wave peak frequency has much better precision. While the results here show the superiority of OPD compared to Parwani, one should note that in a realistic model one would expect even better scale dependence. This is due to the fact that to achieve a first-order phase transition in this model, a very large $g_{12}$ coupling is required to compensate for the small number of degrees of freedom that become massive during the transition. The portal coupling is so large that the uncertainty from matching is significant, despite this being a zero-temperature uncertainty. This is an unfortunate artifact of the toy model we used to develop this analysis. The $\mu$ dependent lines for the thermal parameters are almost in parallel, demonstrating that as we are near the nonperturbative regime where zero-temperature uncertainties are nearly out of control, even if the finite-temperature uncertainties are greatly improved by the OPD resummation scheme. 
Therefore, we expect the gravitational wave prediction to be significantly improved in OPD for more realistic models; furthermore, future work to extend OPD to consistently include RG improvement of the effective potential, and possibly including momentum-dependent terms in the thermal mass resummation, would allow OPD to give more accurate gravitational wave predictions for this particular toy model as well.

\begin{figure}
    \centering
    \includegraphics[width=0.45\textwidth]{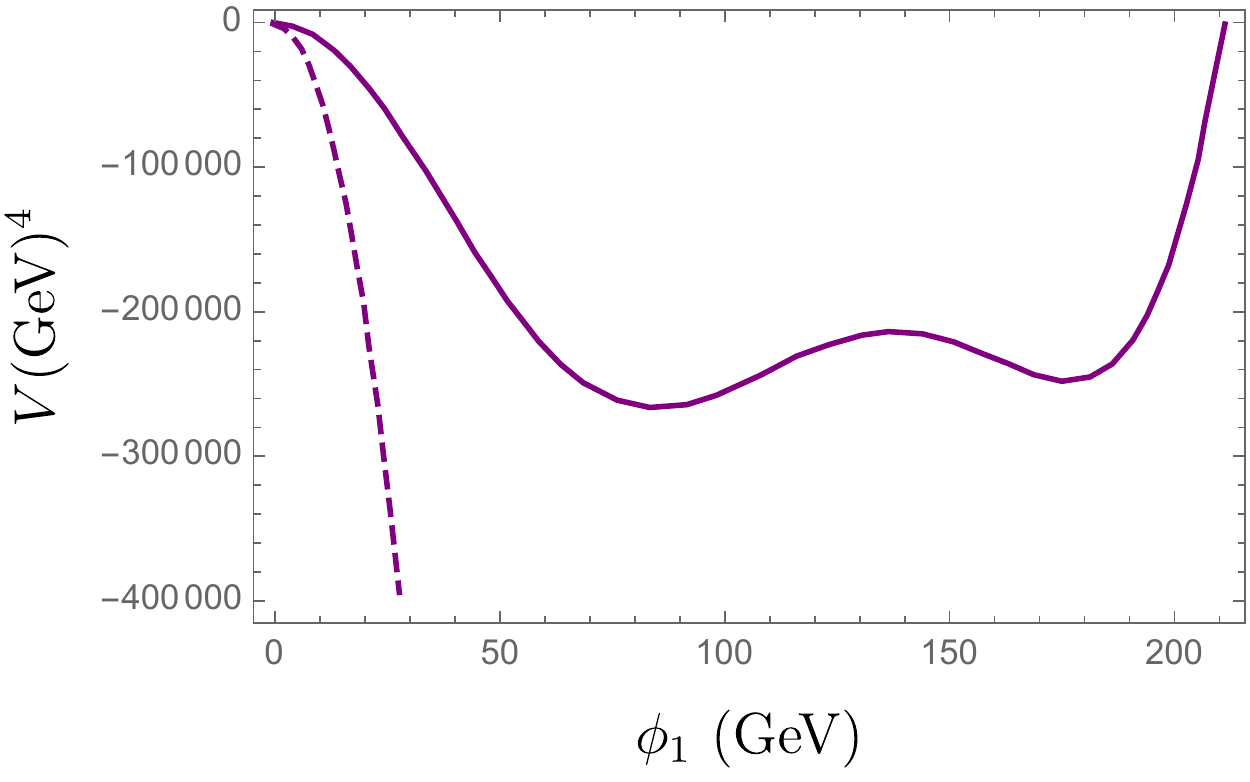}
    \caption{Soft potential in NLO dimensional reduction using the benchmark in Table \ref{Parameter_table} (solid purple) at $T=191.7$ GeV as well as the problematic HT calculation of the two-loop sunset which dominates (dashed purple).}
    \label{fig:frank}
\end{figure}

\begin{figure}
    \centering
    \includegraphics[width=0.45\textwidth]{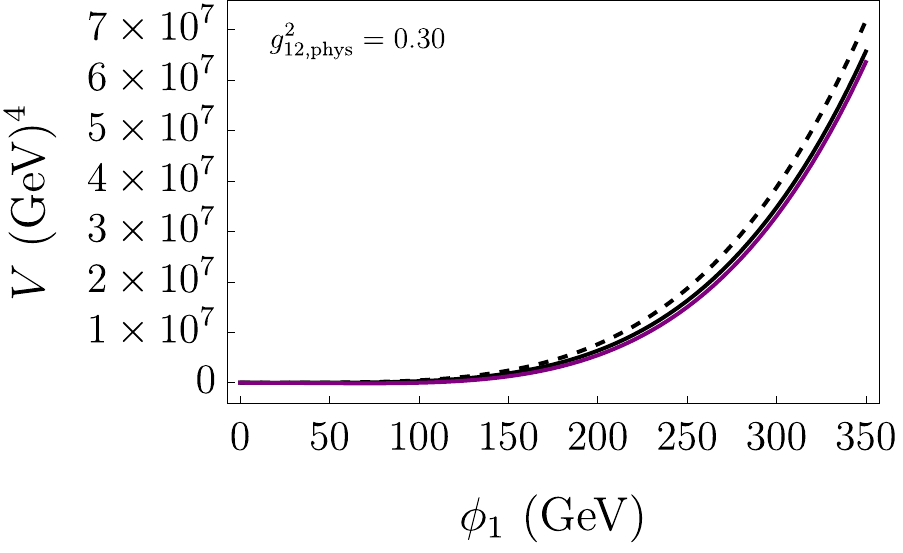}
    \includegraphics[width=0.45\textwidth]{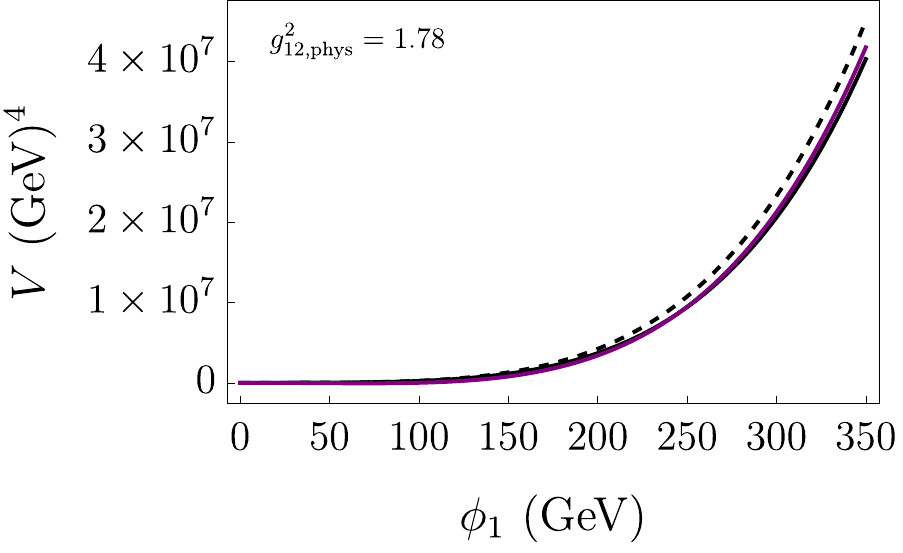}
    \includegraphics[width=0.45\textwidth]{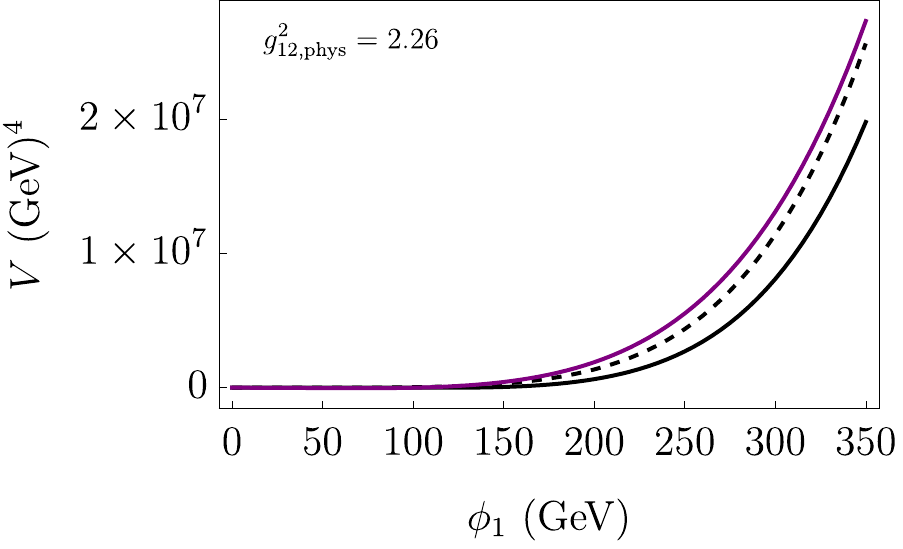}
    \caption{OPD (black) and Parwani (dashed) potentials for a different benchmark from Table.~\ref{Parameter_table}, where the high-$T$ approximation is at least close to  valid, in order to quantitatively compare to DR (purple). $\langle \phi_1 \rangle=400 \, \text{GeV}, M_{1,\text{pole}}=80 \, \text{GeV}, M_{2,\text{pole}}=500 \, \text{GeV}, g_{2,\text{phys}}^{2}=3.60, \mu=500 \, \text{GeV}$ with $g_{12,\text{phys}}^2 = 0.30, 1.78, 2.26$ (top, middle, bottom). Crucially, OPD and DR agree for the first two portal couplings, demonstrating the improvement of OPD compared to Parwani. At higher $g_{12}^2$ (bottom), the high-$T$ approximation used by DR and Parwani breaks down, and only OPD can remain reliable.}
    \label{fig:potcompare}
\end{figure}

Finally, as a last check we should also compare the results to that of dimensional reduction. Unfortunately due to the large couplings involved the high-temperature expansion for $M_2$ badly breaks down near the critical temperature. This makes the use of standard forms of dimensional reduction inappropriate. Particularly problematic is the two-loop sunset diagram that scales as $\phi^2 T^2$ times a large logarithm in the high-temperature expansion (see the last line of equation \ref{eq:nlodr}). This term actually dominates when not Boltzmann suppressed and is responsible for the double barrier behaviour we see in Fig. \ref{fig:frank}. Note the ultrasoft potential does not do better as it is actually unbounded (and contains strange kinks). For this model, the high temperature expansion --- and therefore dimensional reduction --- breaks down for any couplings large enough to catalyze a strong first-order phase transition, so it makes little sense to compare the predicted thermal parameters quantitatively. In this comparison, our calculation gives the best result by default, and while this is physically significant in terms of OPD expanding our capacity to make reliable predictions for phase transitions that are not in the high-temperature regime, it is instructive to consider a case where a quantitative comparison to DR at $O(g_{12,\text{phys}}^4)$ can be made. We therefore compare the potentials directly for differing values of the portal coupling $g_{12}$, without requiring a strong phase transition. We expect that, for small to medium sized $g_{12}$,  OPD and DR should agree with each other and disagree with Parwani; but at some threshold DR and Parwani, which both depend upon the high-temperature expansion, should diverge from OPD, which does not. We find this to be the case, as illustrated by three benchmarks in Fig. \ref{fig:potcompare}. In fact, when the high-temperature expansion breaks down, DR gives a similar prediction to the Parwani calculation. This reinforces the two main points of our analysis: first, as demonstrated analytically in Sec. \ref{Sec:Analytics}, that OPD improves the four-dimensional perturbative calculation to the point where it is of similar accuracy as DR, and second, that OPD improves upon DR by not relying on the high-temperature approximation, which is not respected in many physically relevant phase transitions.

\section{Discussion and conclusion}
\label{Sec:Conclusion}

Understanding the era of cosmological electroweak symmetry is a central question of the next generation of theoretical and experimental efforts. Conventional methods of modelling cosmological electroweak symmetry breaking suffer very difficult theoretical uncertainties. Accurate calculations involving next-to-leading order dimensional reduction are very difficult and, to this day, there exists no global treatment of a BSM scenario involving extra dynamical scalar fields. Further, off the shelf dimensional reduction is in the high-temperature regime, whereas strong phase transitions catalyzed by thermally induced barriers necessitates large couplings where the high-temperature regime is expected to be invalid.

In the context of all of these issues, we are motivated to investigate and further develop the OPD scheme utilizing a recursive solution to the gap equation for scalar masses (though it is worth keeping in mind that the solution of the gap equation can also be used as a replacement for calculating the self energy used in matching relations in the dimensional reduction paradigm as well). OPD includes sizeable contributions neglected by the Parwani scheme, and is therefore required for more accurate calculations. 
In the context of theoretical uncertainties, our results seem to suggest regions for cautious optimism with regard to OPD. The analytic calculation in Sec.~\ref{Sec:Analytics} indicates a similar precision, i.e. scale dependence at fourth coupling order, to dimensional reduction, improving on the second-order dependence of Parwani resummation, while the ease of going beyond the high-temperature approximation in OPD promises additional advantages for phase transitions with sizeable couplings compared to DR. 
Numerical results in Sec.~\ref{Sec:Numerics} for the thermal parameters show an improvement for OPD compared to Parwani, giving further grounds for optimism. All of the thermal parameters except the critical vev display significant reduction in theoretical uncertainty when using OPD. This anomalous behavior for the critical vev does not significantly influence the final gravitational result due to larger dependence on the other thermal parameters, and at any rate we suspect that this is an artifact of the very large couplings necessary for a strong phase transition in our toy model. We also expect this behaviour to improve once a consistent RG-improvement scheme for the OPD  effective potential is implemented.

In summary, both analytical and numerical results clearly indicate the superiority of OPD to the Parwani scheme, which miscounts subleading corrections that are numerically important at large coupling and nonzero scalar vevs, both generally found in phase transitions. Compared to DR, OPD should have parametrically similar theoretical uncertainty once RG-improvement is implemented, with the further advantage of working beyond the high-temperature regime, which is again  ubiquitous for strong first-order phase transition.

Our study represents a first step in a systematic and rigorous development of the OPD scheme for precision high-temperature calculations, and there is a clear itinerary of future directions that must be pursued to apply OPD to realistic extensions of the SM. The highest priority next steps include 
a careful examination of how OPD could be combined with renormalization group improvement of the effective potential to further reduce the theoretical uncertainties; understanding the importance of full momentum dependent self-energy in gap equations; consistent inclusion of gauge bosons in the system of gap equations, and addition of sunset equivalent diagrams in the gauge sector; as well as appropriate modification to the gap equation for multiple symmetry broken fields to handle arbitrary field excursions during the phase transition. Finally, being able to include nonrenormalizable operators would be very useful, given the importance of SMEFT to constrain BSM physics at colliders and elsewhere. 
We are currently working to address these issues and plan to present them in future publications.\\

{\bf Acknowledgments:} We would like to thank Phillipp Schicho for help with DRalgo and Johan L{\"o}fgren for useful comment about the zero-momentum approximation of the self energy in the gap equation.
The research of D. C. was supported in part by a Discovery Grant from the Natural Sciences and Engineering Research Council of Canada, the Canada Research Chair program, the Alfred P. Sloan Foundation, the Ontario Early Researcher Award, and the University of Toronto McLean Award.
The research of J. R. was supported in part by the Natural Science and Engineering Research Council of Canada.
The work of G. W. is supported by World Premier International Research Center Initiative (WPI), MEXT, Japan. G. W. was supported by JSPS KAKENHI
Grant No. JP22K14033.

\appendix 
\section{LOOP-LEVEL MATCHING}
\label{Sec:AppA}

For the matching calculation, we relate the following physical input parameters to the $\overline{\text{MS}}$ Lagrangian parameters at one loop:
\begin{eqnarray}
    (\phi_c,M_{1,\rm{pole}},M_{2,{\rm pole}},g_{12,{\rm phys}},g_{2, {\rm phys}}) \nonumber \\
    \longmapsto (\mu_{1},m_{2},g_{1},g_{12},g_{2})
\end{eqnarray}

For loop-level matching, the standard one-loop renormalization of the tadpole and the self energy diagrams at zero temperature gives three of the conditions relating the physical input parameters to the Lagrangian parameters
\begin{eqnarray}
    V_{1}^\prime (\phi_c) &=& 0 \\
    m_1^{2}(\phi_c) &=& M_{1,\rm{pole}}^2+ \Pi_1 \left (M_{1,\rm{pole}}^2 \right) \\
    m_2^{2}(\phi_c) &=& M_{2,\rm{pole}}^2+ \Pi_2 \left (M_{2,\rm{pole}}^2 \right)
\end{eqnarray}
where $V_{1}$ is the one-loop potential, $m_1^{2}(\phi_c)=-\mu_1 ^2+\frac{1}{2} g_1^2 \phi _c^2$, $m_2^{2}(\phi_c)=m_2^2+\frac{1}{2} g_{12}^2 \phi _c^2$ are the field-dependent masses and the functions can be written in terms of the Lagrangian parameters

\begin{eqnarray}
V_{1}^\prime(\phi_c) &=& -\mu _1^2+\frac{1}{6} g_1^2 \phi _c^2 -\frac{g_1^2 m_1^2(\phi_c)}{32 \pi ^2} \left(\log \frac{\mu^2}{m_1^2(\phi_c)} +1\right) \nonumber \\
&& -\frac{g_{12}^2 m_2^2(\phi_c)}{32 \pi ^2}\left(\log \frac{\mu^2}{m_2^2(\phi_c)} +1\right)
\end{eqnarray}
\begin{widetext}
\begin{eqnarray}
\Pi_{1}\left( p^2 \right) &=& \frac{g_1^2 m_1^2(\phi_c)}{32 \pi ^2}\left(\log \frac{\mu^2}{m_1^2(\phi_c)} +1 \right)+\frac{g_{12}^2 m_2^2(\phi_c)}{32 \pi ^2}\left(\log \frac{\mu^2}{m_2^2(\phi_c)} +1\right) \nonumber \\
&& +\frac{g_1^4 \phi _c^2}{32 \pi ^2} \left(\frac{\sqrt{p^2 \left(p^2-4 m_1^2(\phi_c) \right)}}{p^2}\log \frac{2 m_1^2(\phi_c) +\sqrt{p^2 \left(p^2-4 m_1^2(\phi_c) \right)}-p^2}{2 m_1^2(\phi_c) } +\log \frac{\mu^2}{m_1^2(\phi_c)} +2\right) \nonumber \\
&& +\frac{g_{12}^4 \phi _c^2 }{32 \pi ^2}\left(2+ \log \frac{\mu^2}{m_2^2(\phi_c)} +\frac{\sqrt{p^2 \left(p^2-4 m_2^2(\phi_c)\right)}}{p^2}\log \frac{\sqrt{p^2 \left(p^2-4 m_2^2(\phi_c) \right)}+2 m_2^2(\phi_c) -p^2}{2 m_2^2(\phi_c)} \right)
\end{eqnarray}
\begin{eqnarray}
\Pi_{2}\left( p^2 \right) &=& \frac{g_{12}^2 m_1^2(\phi_c)}{32 \pi ^2}\left(\log \frac{\mu^2}{m_1^2(\phi_c)}+1\right) +\frac{g_2^2 m_2^2(\phi_c)}{32 \pi ^2}\left(\log \frac{\mu^2}{m_2^2(\phi_c)} +1\right) \nonumber \\
&& +\frac{g_{12}^4 \phi _c^2}{16 \pi ^2} \left(-\frac{\left( m_1^2(\phi_c)-m_2^2(\phi_c)+p^2\right)}{2 p^2}\log \frac{m_1^2(\phi_c)}{m_2^2(\phi_c)} +\frac{\sqrt{\left( m_1^2(\phi_c) -m_2^2(\phi_c) \right)^2-2 p^2 \left(m_1^2(\phi_c)+m_2^2(\phi_c)\right)+p^4}}{p^2} \times \right. \nonumber \\
&& \left. \log \frac{\sqrt{\left( m_1^2(\phi_c) -m_2^2(\phi_c) \right)^2-2 p^2 \left( m_1^2(\phi_c) +m_2^2(\phi_c) \right)+p^4}+m_1^2(\phi_c)+m_2^2(\phi_c)-p^2}{2 m_1(\phi_c) m_2(\phi_c)}+\log \frac{\mu^2}{m_2^2(\phi_c)} +2 \right)
\end{eqnarray}
\end{widetext}
Note that for loop-level matching the masses are not obtained by taking the second derivative of the one-loop potential since these correspond to zero external momentum while the pole masses lie at nonzero external momentum. Finally, the remaining two conditions come from the one-loop renormalization of the four point function at zero external momentum. These can be obtained from the derivatives of the one-loop potential which is the tree plus Coleman-Weinberg potential,
\begin{eqnarray}
g_{12, {\rm phys}}^2 &=& \frac{\partial^4  V_1 }{\partial \phi_1^2 \partial \phi_2^2} \Bigg|_{(\phi_1,\phi_2)=(\phi_c,0)} \\
g_{2, {\rm phys}}^2 &=& \frac{\partial^4 V_1 }{\partial \phi_2^4}\Bigg|_{(\phi_1,\phi_2)=(\phi_c,0)}
\end{eqnarray}

\section{TWO-LOOP SUNSET DIAGRAM} \label{Sunset}
The full expression of sunset diagrams used in our results have been calculated in \cite{Parwani:1991gq,PhysRevD.65.045015}
\begin{equation}
V_{\rm sun}= -\frac{g^4 \phi^2}{12}  \left( G_{0}(m^2)+G_{1}(m^2,T)+G_{2}(m^2,T) \right)
\end{equation}
where
\begin{eqnarray}
G_{0}(m^2) &=& -\frac{3 m^2}{2(4\pi)^4} \left( \log^2 \left[ \frac{\mu^2}{m^2}\right] +4 \log \left[ \frac{\mu^2}{m^2}\right] +4 \right. \nonumber \\
&& \left. + \frac{\pi^2}{6} -\frac{8.966523919}{3}\right)
\end{eqnarray}
\begin{eqnarray}
G_{1}(m^2,T) &=& \frac{3}{(4\pi)^2}\left( \log \left[ \frac{\mu^2}{m^2} \right] +2 \right) \int \frac{d^3 q}{(2 \pi)^3} \frac{n_{B}(q)}{\omega_{q}} \nonumber \\
&& + \frac{3}{4(2\pi)^4} \int_0^\infty d q_1 \frac{q_1 n_B(q_1)}{\omega_{q_1}} \nonumber \\
&& \times \int_0^\infty \frac{d q_2}{\omega_{q_2}} \left( q_2 \log \left| \frac{X_+}{X_-} \right| - q_1 \right)
\end{eqnarray}
with
\begin{equation}
    X_{\pm} =\left( \omega_{q_1}+\omega_{q_2}+\omega_{q_1 \pm q_2} \right)^2 \left( -\omega_{q_1}+\omega_{q_2}+\omega_{q_1 \pm q_2} \right)^2
\end{equation}
\begin{eqnarray}
    G_{2}(m^2,T) &=& \frac{3}{4(2\pi)^4} \int_0^\infty d q_1 \frac{q_1 n_B(q_1)}{\omega_{q_1}} 
    \nonumber \\
    && \int_0^\infty d q_2 \frac{q_2 n_B(q_1)}{\omega_{q_2}}  \log \left| \frac{Y_+}{Y_-} \right|
\end{eqnarray}
with
\begin{eqnarray}
    Y_{\pm} &&= \left( \omega_{q_1}+\omega_{q_2}+\omega_{q_1 \pm q_2} \right)^2 \left( -\omega_{q_1}+\omega_{q_2}+\omega_{q_1 \pm q_2} \right)^2 \nonumber \\
    && \times \left( \omega_{q_1}-\omega_{q_2}+\omega_{q_1 \pm q_2} \right)^2 \left( \omega_{q_1}+\omega_{q_2}-\omega_{q_1 \pm q_2} \right)^2
\end{eqnarray}
Here $G_0$ is the pure zero-temperature piece while $G_1, G_2$ are the finite-temperature pieces. The latter terms are included in the 4D perturbative schemes since at high temperature their contributions are the same as one-loop contributions.
\bibliographystyle{apsrev4-1}
\bibliography{references}
\end{document}